\documentclass[wide]{goose-article}

\usepackage{ragged2e}
\usepackage{soul}
\usepackage{xcolor}

\title{How collective asperity detachments nucleate slip at frictional interfaces}

\author[1]{Tom~W.J.~de~Geus}
\author[1]{Marko~Popovi\'{c}}
\author[1]{Wencheng~Ji}
\author[2]{Alberto~Rosso}
\author[1]{Matthieu~Wyart}

\affil[1]{Institute of Physics, \'{E}cole Polytechnique F\'{e}d\'{e}rale de Lausanne (EPFL), Switzerland}

\affil[2]{LPTMS, CNRS, Univ.\ Paris-Sud, Universit\'{e} Paris-Saclay, 91405 Orsay, France
}

\hypersetup{pdfauthor={T.W.J. de Geus}}

\fancypagestyle{firststyle}
{
   \fancyhf{}
   \fancyhead[RO,LE]{\thepage}
   \fancyhead[LO,RE]{Proc.\ Natl.\ Acad.\ Sci., 2019, 116(48):23977-23983, \doi{10.1073/pnas.1906551116}, \arxivid{1904.07635}.}
}

\begin{document}

\twocolumn[
\begin{@twocolumnfalse}

\maketitle

\begin{abstract}
Sliding at a quasi-statically loaded frictional interface can occur via macroscopic slip events, which nucleate locally before propagating as rupture fronts very similar to fracture. We introduce a novel microscopic model of a frictional interface that includes asperity-level disorder, elastic interaction between local slip events, and inertia. For a perfectly flat and homogeneously loaded interface, we find that slip is nucleated by avalanches of asperity detachments of extension larger than a critical radius $A_c$ governed by a Griffith criterion. We find that after slip, the density of asperities at a local distance to yielding $x_\sigma$ presents a pseudo-gap $P(x_\sigma) \sim (x_\sigma)^\theta$, where $\theta$ is a non-universal exponent that depends on the statistics of the disorder. This result makes a link between friction and the plasticity of amorphous materials where a pseudo-gap is also present. For friction, we find that a consequence is that stick-slip is an extremely slowly decaying finite size effect, while the slip nucleation radius $A_c$ diverges as a $\theta$-dependent power law of the system size. We discuss how these predictions can be tested experimentally.

\begin{center}
\textbf{Significance statement}
\end{center}
Understanding how slip at a frictional interface initiates is important for a range of problems including earthquake prediction and precision engineering. The force needed to start sliding a solid object over a flat surface is classically described by a `static friction coefficient': a constant established by measurements. It was recently questioned if such constant exists, as it was shown to be poorly reproducible. We provide a model supporting that it is stochastic even for very large system sizes: sliding is nucleated when, by chance, an avalanche of microscopic detachments reaches a critical radius, beyond which slip becomes unstable and propagates along the interface. It leads to testable predictions on key observables characterising the stability of the interface.

\vspace*{.5em}\noindent\textbf{Keywords:} Friction; Inertia; Avalanches; Fracture; Stick-slip
\end{abstract}

\end{@twocolumnfalse}
]

\sloppy
\thispagestyle{firststyle}

\paragraph{Introduction}

The sliding of a block that rests on a flat surface starts when the applied tangential force passes some threshold $F_S$, which is proportional to the normal force $F_N$. Their ratio defines the friction coefficient $\mu \equiv F_S/F_N$, which typically decreases with increasing sliding velocity when the latter is small \citep{Scholz1976,Baumberger2005,Rabinowicz1956a,Marone1998,Heslot1994}. This phenomenology leads to stick-slip, whereby driving a system quasi-statically results in periods of loading that are punctuated by sudden macroscopic slip events. Experimental observations support that these events proceed by `fracture' \citep{Xia2004,Rubinstein2004,Ben-David2010,Passelegue2013}: after a nucleation phase in which slip appears locally and evolves slowly \citep{Ben-David2011,Ohnaka1990}, a well-defined rupture front appears, that travels ballistically across the frictional interface, unzipping it. This front is accompanied by a stress field in the elastic bulk that is well described by that of a propagating crack \citep{Svetlizky2014,Svetlizky2016}. By contrast, the nucleation phase is much less understood. It is observed that (i) its spatial extension $A_c$ decreases with increasing shear stress \citep{Ben-David2011}, (ii) there is a considerable variability in the tangential force magnitude at which macroscopic slip nucleates \citep{Ben-David2011,Popov2010,Rabinowicz1992} and (iii) acoustic emission \citep{McLaskey2011,Johnson2013} supports that nucleation occurs by bursts of spatially resolvable \cite{McLaskey2011} detachments of micrometer-sized asperities \citep{Bowden1954,Dieterich1994a,Hyun2004}. Explaining these facts is relevant for earthquake predictions \citep{Brace1966} as well as to forecast the variability of the measured friction coefficient \citep{Ben-David2011,Popov2010,Rabinowicz1992}, of importance for precision engineering~\citep{Armstrong-Helouvry1994}.

At a continuum level, rate-and-state models \citep{Dieterich1979,Rice1983,Ruina1983,Scholz1998} are powerful phenomenological descriptions of frictional interfaces, in which friction depends on the sliding velocity as well as some history-dependent state $\phi$ of the interface. Several length scales appear in these approaches \cite{Ruina1983,Ohnaka1990}, including a Griffith length beyond which sustained slip-pulses can propagate, as well as a larger length at which these pulses can nucleate fracture, reminiscent of a first-order phase transition\footnote{In this dynamical phase transition, the order parameter is the strain rate while the control parameter is the stress. A first-order transition therefore refers to a discontinuous strain rate \emph{vs} stress behaviour.} \citep{Brener2018}. Yet these descriptions are coarse-grained and phenomenological, and connecting them with asperity-level phenomena where disorder, that arises from surface roughness, is preponderant remains a challenge. Likewise, the precise meaning of the state variable $\phi$, often thought as capturing the ageing of contacts, remains to be clarified. One microscopic view is that a sliding frictional interface shares similarities with the plastic flow of amorphous materials \cite{Baumberger1999}. Interestingly, it was recently observed that the `state' of bulk amorphous materials \citep{Lemaitre07,Karmakar10a} can be quantified by the density of soft spots about to yield locally, which scales as $P(x_\sigma) \sim (x_\sigma)^\theta$, where $x_\sigma$ is the stress increment required for a given spot to yield\footnote{The scaling $P(x_\sigma) \sim (x_\sigma)^\theta$ holds only for small $x_\sigma$, namely for the soft spots. The term ``pseudo-gap'' refers an exponent $\theta>0$ that corresponds to a singular depletion for small $x_\sigma$ \cite{Muller14}.}. The existence of a non-trivial exponent $\theta>0$ was shown to be a necessary consequence of the long-range elastic interactions and of the non-monotonic (varying in sign) stress redistribution triggered by plastic events \cite{Lin14a,Lin14,Lin15,Lin16}. This parallel raises the intriguing possibility that frictional interfaces are characterised by some exponent $\theta$ as well.

Our goal is to propose a description of frictional interfaces that captures disorder at the asperity level, long-range elastic interactions between local slip events, and inertia. When the latter is absent, the physics is well understood and falls into the universality class of the depinning transition\footnote{The depinning transition occurs for example when an elastic manifold is pulled through a disordered medium \citep{Fisher1998}.} with monotonic interactions \citep{Fisher1998, Kardar1998, Ferrero2013}, for which a continuous transition at a unique, well-defined, macroscopic critical force $F_c$ \citep{Middleton1992} separates a flowing and an arrested phase. There exists no macroscopic stick-slip, and at $F_c$ motion corresponds to power law avalanches in which many local slip events act in concert. However, what happens to this scenario when inertia matters (as it does at a frictional interface) is a matter of debate. The popular view is that inertia destroys criticality: the transition becomes first order, stick-slip appears, and for significant inertia macroscopic slip events are nucleated by a few asperities acting together \citep{Fisher1997, Dahmen1998}. However, another scenario has been proposed by Schwartz et al.\ \citep{Schwarz2003} based on a simplified cellular automaton model describing short-range elasticity, in which stick-slip is a slowly decaying finite size effect that vanishes in the thermodynamic limit, for which the transition is continuous. Nevertheless Schwartz et al.\ later argued \citep{Maimon2004} that for physical systems the scenario developed in \citep{Fisher1997} was presumably correct, and that the conclusions of \citep{Schwarz2003} on the absence of hysteresis in infinite systems were non-generic and only valid for a finely tuned model.

In this work we introduce a novel numerical model of flat, homogeneously loaded, frictional interfaces where inertia is properly treated by discretising the bodies in contact by finite elements. Asperities at the interface are described by elements endowed with a random potential that represents the presence of surface roughness, allowing for sudden local slip events when a local (random) threshold stress is reached\footnote{This abstraction presents similarities to the treatment of the local rearrangement of particles in amorphous solids as shear-transformation zones \citep{Argon79}, and allows similar numerical treatment \citep{Homer2009, Jagla07, Jagla2017}.}. Our model thereby differs from existing ones as no microscopic constitutive friction model (such as slip-weakening, velocity-weakening, or rate-and-state; see e.g.\ the spring--block models of \citep{Andrews1976, Tromborg2014, Tromborg2015, Amon2017} and references therein) is presumed. In contrast, slip-weakening emerges as a consequence of mechanical noise (elastic waves generated by inertia) emitted when asperities detach, which can cause the nucleation of macroscopic slip.

Our main findings are that: (a) surprisingly, the scenario developed in \citep{Schwarz2003} is correct: stick-slip is a finite-size effect, although we find its power law decay with system size to be so slow that it is significant even in very large systems. In that regime, the friction coefficient is intrinsically stochastic, consistent with (ii) above. (b) Despite the interaction being monotonic, due to inertia the interface presents a non-trivial exponent $\theta$ characterising a pseudo-gap in the density of asperities about to yield. We argue that this conclusion will hold more generally to depinning problems with inertia and long-range elasticity. (c) Nucleation is triggered when avalanches get bigger than a critical radius, governed by a Griffith criterion, that diverges as a power law of the system size. Experimentally, the presence of avalanches is consistent with the measured distribution of acoustic emission (iii), while Griffith's criterion is supported by a decreasing nucleation length with increasing stress (i). We relate the exponents associated with properties (a) and (c) to $\theta$ and those characterising avalanches, and confirm our predictions numerically. Finally, we propose experiments to test our results and measure $\theta$.

\paragraph{Model}

\begin{figure}
  \centering
  \includegraphics[width=\linewidth]{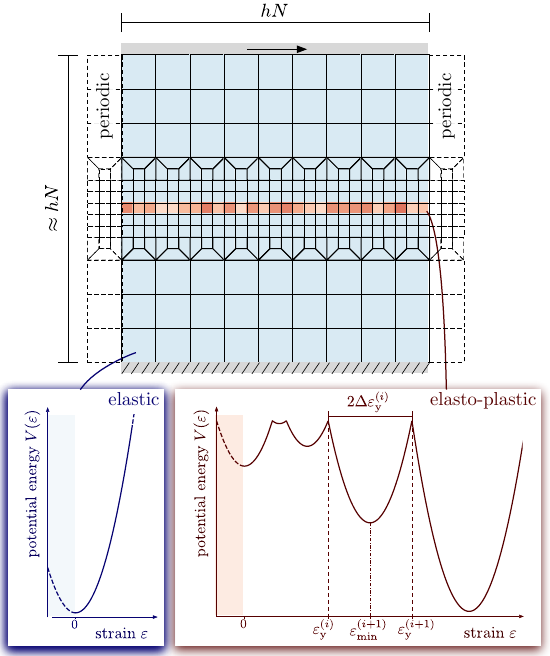}
  \caption{Finite element model of a frictional interface. The two bulk solids display linear elasticity, corresponding to a purely quadratic potential $V(\varepsilon)$ in each blue element (where the shear strain $\varepsilon$ is the norm of the deviatoric part of the strain tensor). The weak middle layer represents the detachment and reattachment of asperities. The potential in each such element (referred to as ``block'', in red) is random, and made of sequences of parabola separated by cusps, leading to sharp plastic events. The system is subjected to quasi-static simple shear by keeping the bottom boundary fixed while displacing the top boundary using infinitesimal strain increments after which energy is minimised.}
  \label{fig:model}
\end{figure}

The geometry of our set-up is illustrated in Fig.~\ref{fig:model}. It comprises a frictional interface (in red) embedded between two identical isotropic linear elastic materials (in blue), all discretised using finite elements and interacting in the same manner. The elements along the frictional interface (referred to as ``blocks'') are plastic: they respond elastically (with the same elastic constants as the elastic bodies) up to a local yield strain (see below). To mimic energy leakage at the boundaries (by the transmission of elastic waves), we consider a viscous damping in the bulk whose magnitude is such that waves travel on the order of the system size before decaying (see \emph{Methods}). The system is periodic in the horizontal direction, while the top and bottom boundaries are used to impose an event-driven quasi-static simple shear. In this protocol, the strain is increased up to the next plastic event (the response to this increase is purely elastic and in mechanical equilibrium), after which an infinitesimal strain increment is applied, triggering (an avalanche of) plasticity. Once motion stops, this sequence is repeated.

The frictional interface consists of $N$ `elasto-plastic' blocks (finite elements) of linear size $h$, each representing one or a few asperities\footnote{More specifically, each block corresponds to the so-called Larkin length \citep{Cao2018} below which asperities always collectively rearrange. Our predictions below apply if the Larkin length is much smaller than the whole system size. In the experiments of \citep{Ben-David2011,Svetlizky2016,Svetlizky2014,Passelegue2013,Ben-David2010,Rubinstein2004}, the nucleation length is found to be quite smaller than the system size, consistent with this assumption. See Appendix~\ref{sec:statistics} for quantitative statements.}. Similar blocks are used in models of plasticity of amorphous materials \citep{Jagla07, Jagla2017}. Each block is characterised by a random potential $V(\varepsilon)$ function of the equivalent shear strain $\varepsilon$ (the norm of the deviatoric part of the strain tensor), see Fig.~\ref{fig:model}(bottom). $V(\varepsilon)$ is constructed from a sequence of quadratic potentials of identical curvature, whose intersections define the yield strains. Disorder is introduced by randomly drawing the yield strains $\Delta\varepsilon_\mathrm{y}$ from some distribution, independently for each block. We chose a Weibull distribution $P(\Delta\varepsilon_\mathrm{y}) = k \; (\Delta\varepsilon_\mathrm{y})^{k-1} \exp\left[ -(\Delta\varepsilon_\mathrm{y})^k \right]$
with $k = 2$. To acquire statistics we consider an ensemble of independent realisations and focus on the two-dimensional case where larger systems can be reached (see \emph{Methods} and Appendix~\ref{sec:model} for details).

\paragraph{A single plastic event}

Under shear loading, a block responds linearly up to reaching the local yield strain, corresponding to a cusp in $V(\varepsilon)$. Passed that point, the block releases some of its elastic energy, and settles in a new equilibrium position determined by the potential energy of the element and the interaction with its surroundings. Such plastic shear strain leads to a permanent redistribution of shear stress in the system, that decays as a force dipole $1/r^d$ \citep[][and Appendix~\ref{sec:response}]{Eshelby1956}, with $r$ being the distance from the block and $d$ the dimension of the space (here $d=2$). Along the weak layer the kick in shear stress is strictly positive (and decays in space as $1/r^d$), corresponding to a monotonic interaction. This effect alone can destabilise other blocks, leading an avalanche of yielding events.

In addition, each yielding event emits elastic waves, causing a transient stress, whose amplitudes decay in space as a force monopole $1/r^{d-1}$. This effect can trigger yielding of blocks that would have remained stable otherwise, causing a dynamical weakening effect: plastic activity leads to more inertial mechanical noise, which in turn creates more plastic activity.

\paragraph{Avalanches as precursors of macroscopic slips}

\begin{figure}
  \centering
  \includegraphics[width=\linewidth]{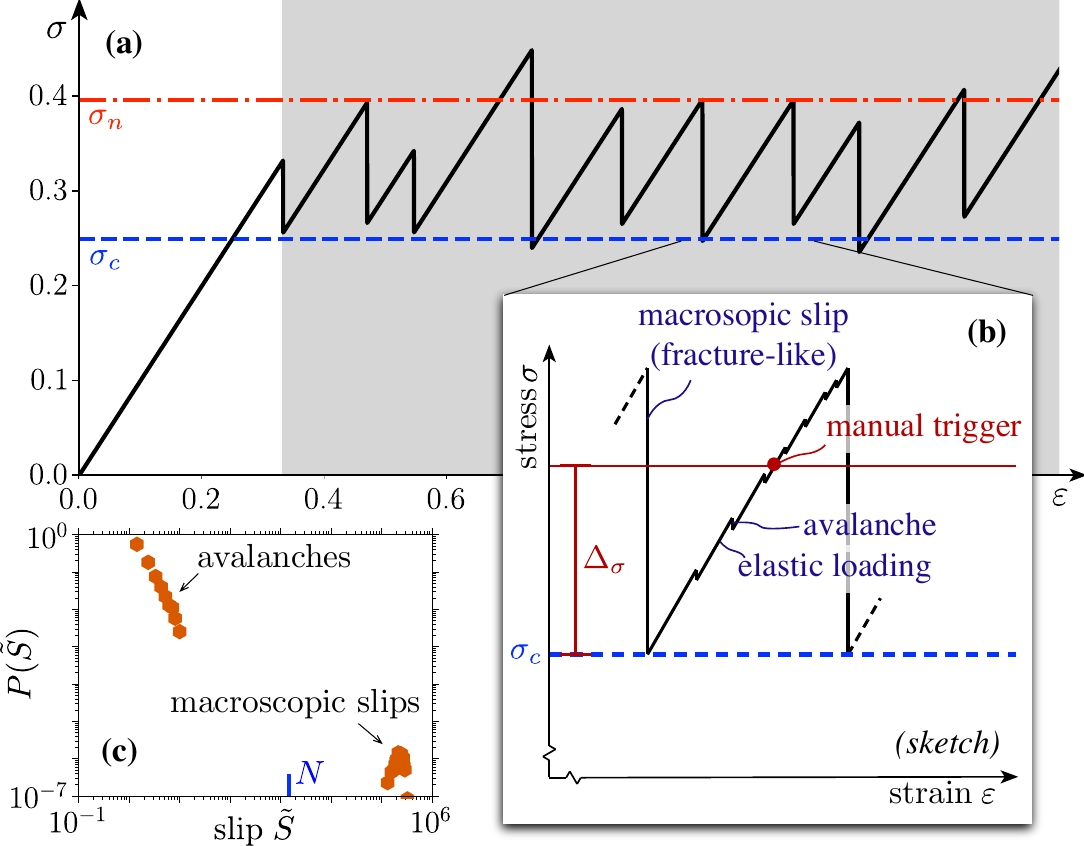}
  \caption{\textbf{(a)} Example of the stress--strain, $\sigma$-$\varepsilon$, response of a system of size $N = 3^6$. The considered steady state is highlighted in grey. The ensemble average stress before and after macroscopic slip (a system-spanning event), $\sigma_n$ and $\sigma_c$, are indicated using dashed lines. \textbf{(b)} Schematic of a stick-slip cycle. The stress $\Delta_\sigma \equiv \sigma - \sigma_c$ at which events are manually triggered to gather statistics, is also indicated. Note that the apparent absence of small stress drops due to avalanches in (a) is because these avalanches are rare, as detailed in the text. \textbf{(c)} Distribution of slip size, $P(\tilde{S})$, where $\tilde{S}$ is defined as the total number of times the blocks yielded during a single event. Note that we present all data, in (c) and in Figs.~\ref{fig:push}--\ref{fig:theta}, in terms of the largest system ($N = 3^6 \times 2$), and validate their robustness in terms of system size in Appendix~\ref{sec:robustness}.}
  \label{fig:typical}
\end{figure}

A typical stress--strain response is shown in Fig.~\ref{fig:typical}(a). The system first responds elastically, followed by a steady state stick-slip behaviour (highlighted in grey). The stick-slip phase consists of loading intervals punctuated by macroscopic slip during which all blocks yield many times, on average causing the stress to drop from $\sigma_n$ to $\sigma_c$ (see Fig.~\ref{fig:typical}(a)). Such macroscopic slips are fracture-like, as supported by the time evolution that presents a ballistic propagation front, that travels at a super-shear velocity, consistent with the recent experiments \citep{Rubinstein2004,Ben-David2010,Svetlizky2014,Svetlizky2016}, see Appendix~\ref{sec:response}.

As in experiments \citep{Johnson2013}, we observe microscopic activity during the loading phases. It corresponds to events that failed to nucleate macroscopic slip, and as such are important to analyse. The distribution of slip sizes $\tilde{S}$, defined as the total number of times that blocks yield during an event shows a clear separation in two types of events: \emph{macroscopic slips} at $\tilde{S} \gg N$ (indicating that blocks have yielded many times), and \emph{avalanches} that occur during the loading phase\footnote{\emph{Macroscopic slips} are events in which all blocks yield at least once, and \emph{avalanches} are all other, localised, events.}; see Fig.~\ref{fig:typical}(c) and the sketch in Fig.~\ref{fig:typical}(b). However, the occurrence of avalanches is too rare to be insightful.

To gain more information about the avalanches, we manually trigger events at different stresses $\Delta_\sigma \equiv \sigma - \sigma_c$, by locally applying a shear displacement perturbation to a randomly selected block along the weak layer (see Appendix~\ref{sec:model}). If all blocks were elastic, the displacement would simply snap back to the original equilibrium configuration. But for the elasto-plastic blocks an avalanche can be triggered, leading to a new equilibrium state.

\begin{figure}
  \centering
  \includegraphics[width=.95\linewidth]{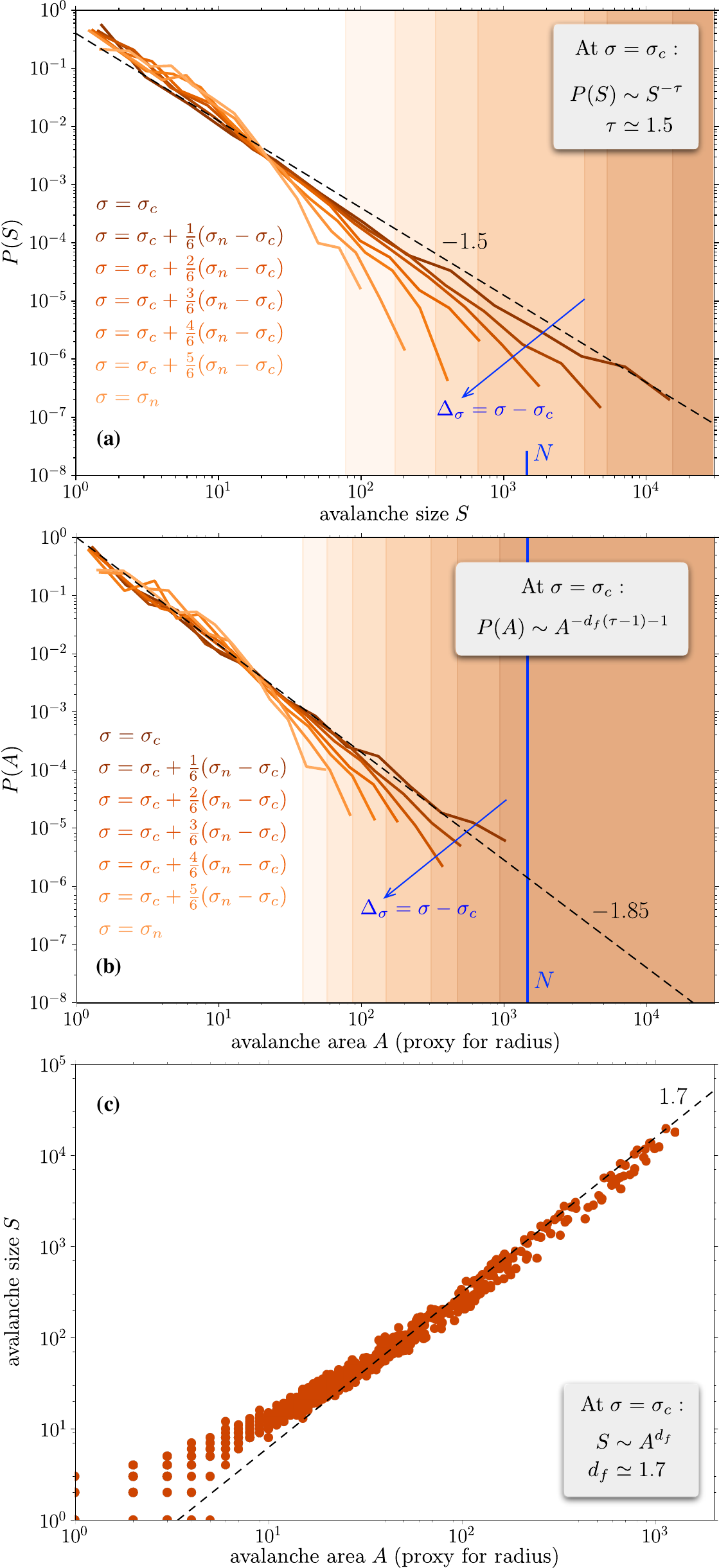}
  \caption{\textbf{(a)} Distribution of avalanche sizes, $P(S)$, at different stresses $\sigma$ at or above the critical stress $\sigma_c$. Note that macroscopic slips, whereby the avalanche grew unstable and spanned the entire system, have been filtered from this distribution. The highlighted regions correspond to the cutoff size, $S_c$ (see text, below Eq.~\eqref{eq:griffith}, for measurement), for each of the shown stresses. \textbf{(b)} Same as (a) for the area $A$ of avalanches. \textbf{(c)} Measurement of the fractal dimension, $d_f$: the relationship between the area, $A$, and the avalanche size, $S$, at $\sigma = \sigma_c$. The expected scaling and measured exponents have been included in text boxes. In all figures the dashed line marks the power law scaling with the indicated exponent.}
  \label{fig:push}
\end{figure}

We first focus on $\Delta_\sigma=0$. The distribution of avalanches sizes, $P(S)$, is obtained by eliminating events that result in macroscopic slip (defined as an event in which all $N$ blocks yielded at least once). Strikingly, we find a power law distribution of avalanches at $\sigma = \sigma_c$:
\begin{equation}
  P(S) \sim S^{-\tau}
  \label{eq:rho(S)}
\end{equation}
with the exponent $\tau \simeq 1.5$ (Fig.~\ref{fig:push}(a)) and a fractal dimension $d_f \simeq 1.7$. The latter relates spatial extension $A$ (the number of sites that yielded at least once) of an avalanche to its size:
\begin{equation}
  S \sim A^{d_f}
  \label{eq:df}
\end{equation}
(Fig.~\ref{fig:push}(c)). These results imply
\begin{equation}
  P(A) \sim A^{-d_f(\tau-1)-1}
  \label{eq:rho(A)}
\end{equation}
as confirmed in Fig.~\ref{fig:push}(b). We conclude that the stress $\sigma_c$, after macroscopic slip, is a critical point at which the distribution of avalanche sizes is scale free.

\paragraph{Mechanism for `fracture' nucleation}

Our central observation in Figs.~\ref{fig:push}(a,b) is that increasing $\sigma$ above the critical point $\sigma_c$ leads to a smaller and smaller cutoff $A_c$ and $S_c$ for the distribution $P(A)$ and $P(S)$. At first glance this is surprising, since at large stresses one may expect avalanches to be bigger. In fact, this cutoff signifies that large avalanches run away, and lead to macroscopic slip (not included in these distributions). Thus the cutoff $A_c$ and $S_c$ characterise the size of the avalanches required to nucleate a macroscopic slip event.

\begin{figure}
  \centering
  \includegraphics[width=\linewidth]{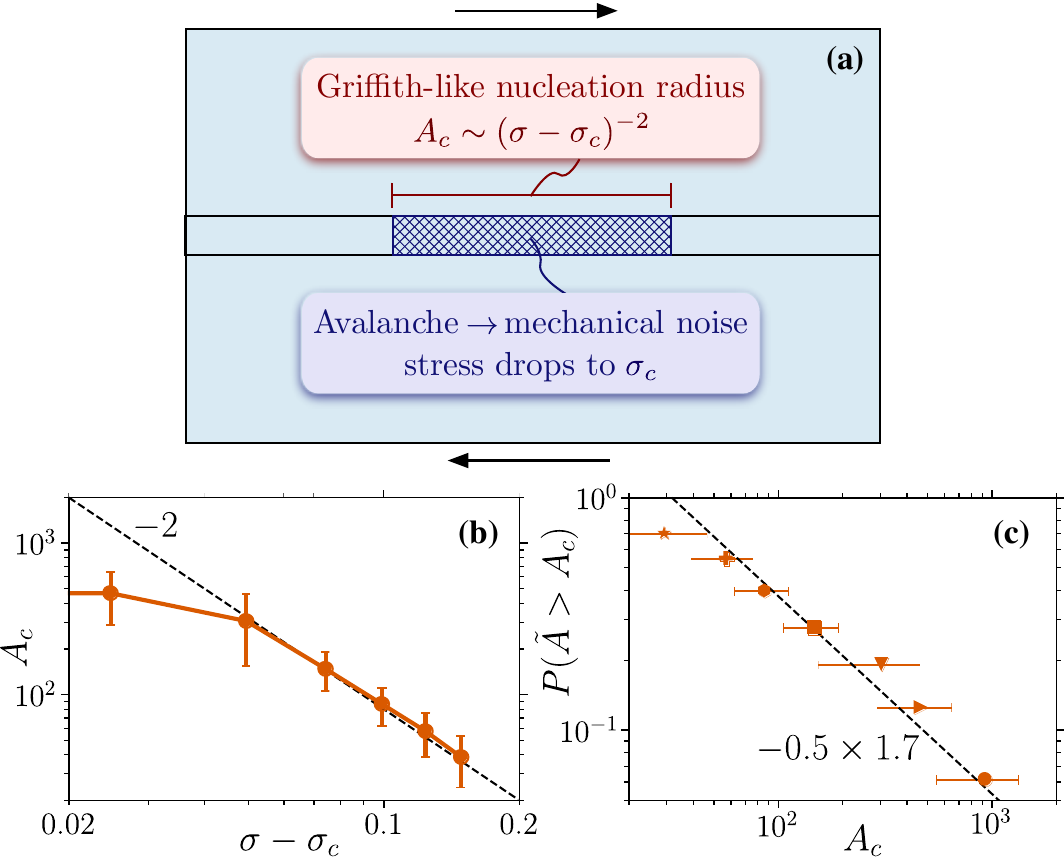}
  \caption{\textbf{(a)} Sketch of nucleation of fracture-like macroscopic slip by an avalanche. \textbf{(b)} Scaling of $A_c$ as predicted by the Griffith-like criterion for nucleation of macroscopic slip in Eq.~\eqref{eq:griffith}. \textbf{(c)} The probability that an event (an avalanche or a macroscopic slip) has a radius larger than $A_c$, as predicted by Eq.~\eqref{eq:P(Ac)}. The dashed line in (b,c) marks the power law scaling with the indicated exponent.}
  \label{fig:push_scaling_A}
\end{figure}

We now propose a scaling relationship for $A_c$ as a function of $\sigma-\sigma_c$. We posit that $\sigma_c$ is the maximum stress that the frictional layer can locally carry in the presence of endogenous inertial mechanical noise. This noise is generated by the ballistic pulses of stress emitted by failing blocks when the interface is in the process of plastically rearranging locally. Now consider triggering an avalanche at $\sigma>\sigma_c$. Avalanches are compact objects (as $d_f>1$), implying that each block yields many times, inducing a large inertial mechanical noise. On average, this will reduce the stress inside the avalanche to $\sigma_c$ (see sketch of Fig.~\ref{fig:push_scaling_A}(a)), while at large distances from the avalanche the stress remains $\sigma > \sigma_c$. This mismatch leads to stress concentrations at avalanche's edges proportional to a stress intensity factor $(\sigma - \sigma_c)\sqrt{A}$. As postulated by Griffith \citep{Anderson05,Griffith1921}, a fracture instability\footnote{Note that in contrast to an opening crack, which cannot carry any stress, a stress $\sigma_c$ can still be carried during macroscopic slip.} will take place when the intensity factor reaches a threshold, implying:
\begin{equation}
  A_c \sim (\sigma - \sigma_c)^{-2}
  \label{eq:griffith}
\end{equation}
(for any $d$). We confirm this result in Fig.~\ref{fig:push_scaling_A}(b), supporting our hypothesis that the stress inside the avalanche on average drops to $\sigma_c$. The departure from scaling in Fig.~\ref{fig:push_scaling_A}(b) at small $\Delta_\sigma$ is due to the value of $A_c$ being so large that our measurements suffer from finite size effects, see Appendix~\ref{sec:robustness}. Note that we measure $A_c$ using the ratio of successive moments to extract $A_c \equiv \langle A^{p+1} \rangle / \langle A^{p} \rangle$. In practice we use $p = 4$ to be more sensitive to the biggest avalanches while still having good statistics, but our results are robust to different choices, see Appendix~\ref{sec:robustness}.

\paragraph{Macroscopic slip}

Macroscopic slip nucleates when, by chance, an avalanche exceeds the nucleation radius $A_c$. As stress increases, more and more avalanches are triggered, while concurrently the nucleation radius shrinks. Nucleation of macroscopic slip thus becomes more and more likely with increasing stress. Typically, macroscopic slip will have happened when the stress is sufficiently large such that
\begin{equation}
  n_a \, P(A > A_c) \sim 1
  \label{eq:nucleation}
\end{equation}
where $n_a$ is the number of triggered avalanches that have occurred as a result of a stress increment $\Delta_\sigma = \sigma - \sigma_c$, and ${P(A > A_c)}$ is the fraction of those avalanches that exceed the radius at which macroscopic slip is nucleated. Eq.~\eqref{eq:nucleation} thus sets the typical value of stress, $\sigma_n$, at which macroscopic slip occurs.

The probability that an avalanche has a radius larger than $A_c$ follows from Eqs.~(\ref{eq:rho(A)},\ref{eq:griffith}):
\begin{equation}
  P(A > A_c) \sim A_c^{d_f (1-\tau)} \sim (\sigma - \sigma_c)^{-2 d_f (1-\tau)}
  \label{eq:P(Ac)}
\end{equation}
as verified in Fig.~\ref{fig:push_scaling_A}(c).

\begin{figure}[htp]
  \centering
  \includegraphics[width=\linewidth]{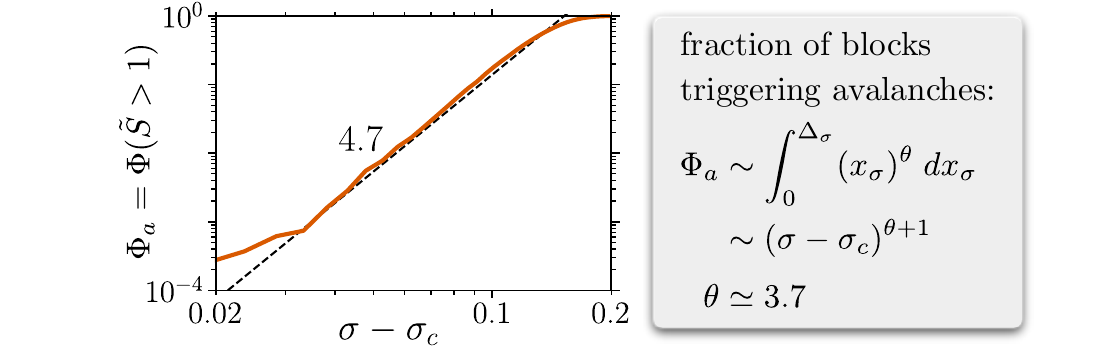}
  \caption{Cumulative probability density of avalanches as a function of the stress increase $\Delta_\sigma = \sigma - \sigma_c$, measured by counting the number of avalanches during quasi-static loading. It provides a direct measurement of $P(x_\sigma) \sim (x_\sigma)^\theta$ characterising the interface stability. The dashed line marks the power law scaling with the indicated exponent.}
  \label{fig:theta}
\end{figure}

The number of avalanches $n_a$ follows from the distribution $P(x_\sigma)$ of the stress increment $x_\sigma$ required for a given block to yield for the first time after a big slip event and trigger an avalanche. Let us assert for the moment (and confirm below) that this distribution follows a power law:
\begin{equation}
  P(x_\sigma) \sim (x_\sigma)^\theta
  \label{eq:rho(x)}
\end{equation}
In that case, the fraction of blocks that triggers an avalanche upon increasing the stress by $\Delta_\sigma = \sigma - \sigma_c$ scales like
\begin{equation}
  \Phi_a \sim \int_0^{\Delta_\sigma} (x_\sigma)^\theta \;d x_\sigma \sim (\sigma-\sigma_c)^{\theta + 1}
  \label{eq:Phi_a}
\end{equation}
This allows us to measure $\theta$ by counting the number of avalanches during the loading periods. We find a non-trivial exponent $\theta \simeq 3.7$, as shown in Fig.~\ref{fig:theta}. For the number of avalanches, we thus get:
\begin{equation}
  n_a = N \Phi_a \sim N (\sigma - \sigma_c)^{\theta + 1}
  \label{eq:n_a}
\end{equation}

Inserting Eqs.~(\ref{eq:P(Ac)},\ref{eq:n_a}) into Eq.~\eqref{eq:nucleation} leads to:
\begin{equation}
  \sigma_n - \sigma_c
  \sim N^{\frac{-1}{2 d_f (\tau-1) + \theta + 1}}
  \sim N^{-0.16}
  \label{eq:scaling:delta_sigma}
\end{equation}
It follows from this argument that: (i) The stress $\sigma_n$ at which macroscopic slip nucleates is stochastic, as embodied by Eq.~\eqref{eq:P(Ac)}. (ii) The stick-slip amplitude $\sigma_n - \sigma_c$ eventually vanishes as the number of asperities $N \rightarrow \infty$. Stick-slip is thus a finite size effect, yet the decay is so slow that it is expected to persist in realistic systems. In a truly infinite system avalanches should be power law distributed. (iii) The fracture nucleation radius diverges as:
\begin{equation}
  A_c (\sigma = \sigma_n)
  \sim N^{\frac{2}{2 d_f (\tau-1) + \theta + 1}}
  \sim N^{0.32}
  \label{eq:scaling:Ac}
\end{equation}

\paragraph{Argument for pseudo-gap $P(x_\sigma) \sim (x_\sigma)^\theta$}

The stability distribution, $P(x_\sigma)$, can in general obey one of three scenarios at small $x_\sigma$: (i) \emph{Depinning:} a finite number of blocks can yield after a small increase of stress, characterised by an exponent $\theta = 0$ \citep{Lin14a,Fisher1998}. (ii) \emph{A pseudo-gap:} the number of blocks that can yield vanishes only at $x_\sigma = 0$, i.e.\ $\theta > 0$. (iii) \emph{Gap:} a small depleted region at small $x_\sigma$, such that $P(x_\sigma < D) = 0$ for some small but finite $D$, thus requiring a finite increase of stress to destabilise any block. This scenario appears to be required to get true stick-slip as ${N\rightarrow\infty}$.

Our data in Fig.~\ref{fig:theta} and other measurements below support scenario (ii). We now exclude the depinning scenario based on a stability argument. In the presence of inertia, the temporary stress overshoot can destabilise blocks that would otherwise stop at small $x_\sigma$. Stability of the system requires that the number of blocks that are destabilised by one event does not diverge when the system size goes to infinity. This leads to the condition $\theta > 0$ as follows: When a block fails, it emits a temporary stress overshoot $\sigma_I \sim 1/r$ (in 2D). The probability that this will destabilise other blocks is $P(x_\sigma < \sigma_I ) \sim r^{-(\theta + 1)}$. Consequently, in a system of size $R$ the number of destabilised blocks $n_f \sim \int_h^R r^{-(\theta + 1)} dr$ diverges as $R \to \infty$, unless $\theta > 0$ (see Appendix~\ref{sec:stability} for a more general argument).

We currently do not have a theory for the value of exponent $\theta$, but preliminary observations indicate that $\theta$ is non-universal. Building a theory to understand $\theta$ should explain the following observations (presented in detail in Appendix~\ref{sec:stability}): (a) The blocks for which $x_\sigma$ is very small following a macroscopic slip event typically lie in a shallow well followed by another shallow well in the block \citep{DeGeus2015a}. (b) As a consequence, when triggered they tend to lead to small slips, and are less likely to trigger slip in other sites. As a result, there exists another exponent $\theta' \simeq 2.5 \leq \theta$ characterising the density of sites at a distance $x_\sigma$ to yield, unconditioned to subsequently triggering an avalanche (our argument and measure in Fig.~\ref{fig:theta} is conditioned to sites triggering an avalanche). (c) The exponent $\theta$ is not universal and depends on the specific choice of disorder, in particular on the parameter $k$ entering the Weibull distribution, and characterising the probability to find narrow wells. Using $k=1.2$ instead of $k=2$, we find $\theta' \simeq 1.4$.

The presence of the strong depletion of the number of the almost unstable blocks, induced by the macroscopic slips, ($\theta > 0$ for Eq.~\eqref{eq:rho(x)}) can be a possible explanation of the observed exponent $\tau \simeq 1.5$ that characterises the distribution of the avalanche sizes in Eq.~\eqref{eq:rho(S)}. For the depinning transition, in the overdamped limit, we know that $\tau_{\text{dep}}=1.28$ \citep{Bonamy2011,Moulinet2004}. However, we also know that each such avalanche is a collection of spatially disconnected slipping regions, called clusters. When treated as separate events, the distribution of avalanche sizes of individual clusters is also scale free, but with a larger exponent, $\tau_{\text{clus}} \simeq 1.56$ \cite{Laurson2010}, close to our measured $\tau \simeq 1.5$. The existence of these disconnected clusters is a consequence of the long-range nature of the elastic interactions \cite{Joanny1984}: a slipping block is a source of instability for the neighbourhood, but also for blocks far away that are very close to their yield stress (have a small $x_\sigma$). The presently observed strong depletion of $P(x_\sigma)$ for small $x_\sigma$ implies that there are very few blocks close to yielding, thus reducing the likeliness of triggering a `secondary', disconnected, avalanche.

\paragraph{Discussion}

We have introduced a model of a frictional interface that includes microscopic disorder at the asperity scale, long-range elastic coupling between local slip events, and the propagation of inertial waves. Our results support a description unifying collective avalanches of asperity detachments and fracture-like macroscopic slip events, in which the former nucleates the latter once a critical avalanche size is reached. These predictions are compatible with existing observations: the presence of avalanches is consistent with the measured distribution of acoustic emission \citep{McLaskey2011,Johnson2013}, while Griffith’s criterion is supported by a decreasing nucleation length with increasing stress \citep{Ben-David2011}. Two surprises emerge from our predictions. First, a key aspect of the interface is the distribution of asperities about to yield, which is very much depleted and characterised by a non-trivial exponent $\theta$ after a macroscopic slip event. Second, we find that the transition to sliding is a continuous transition in the thermodynamic limit, but that finite size effects decay extremely slowly: the stress drop is a stochastic quantity whose typical scale decays as $N^{-0.16}$ and will thus persist in very large system, leading to a slowly diverging nucleation radius $A_c\sim N^{0.32}$.

Our predictions could be quantitatively tested in nearly flat and homogeneously loaded samples, which may be achievable experimentally using the apparatus of~\citep{Sahli2018, Bureau2000}. In particular, microscopic slip events could be measured using (an array of) mechanical or acoustic sensors like in~\citep{McLaskey2011, Rubinstein2004}. Their cumulative number while quasi-statically loading the sample by a stress increment $\Delta_\sigma$ after a macroscopic slip event is proportional to $(\Delta_\sigma)^{\theta + 1}$, thus allowing one to access empirically the pseudo-gap exponent $\theta$. Moreover, well-separated avalanches could be acquired using our trick of triggering avalanches at different stress levels after macroscopic slip, for instance by supplying a focused acoustic signal to the system and measuring the magnitude of the mechanical or acoustic response. We expect the distribution of the magnitude to display a power law $P(S) \sim S^{-\tau}$ with a cutoff $S_c$ decreasing as $S_c \sim A_c^{d_f} \sim (\Delta_\sigma)^{-2 d_f}$. Beyond $\tau$, such a measurement would thus also yield an estimate of the fractal dimension of the avalanches $d_f$, without the need to spatially resolve the avalanches. As a reference, we document the statistics needed to extract these exponents reliably using our model in Appendix~\ref{sec:statistics}.

There is an apparent opposition between the description presented here, and rate-and-state models where velocity-weakening is assumed to hold in the continuous limit, and nucleation stems from a first order transition. It would be very interesting to study how these two scenarios evolve when disorder is present at all scales (including the fact that the surface can have a roughness exponent, and the loading can be very heterogeneous). It is possible that in our approach as well, the transition becomes first order for certain statistics of the disorder. We view it as an important extension of the present work. Another important extension is the inclusion of creep. It may be readily achievable by putting our model in contact with a thermal bath, since in that case individual asperities will age to find a deeper nearby well.

Finally, it is interesting to ask which class of dynamical transitions can become first order due to inertia, and which cannot. The role of inertia has been studied recently in amorphous materials \citep{Karimi2017,Nicolas2016,Salerno2013,DeGiuli17a,Vasisht2018}, where it leads to a large pseudo-gap exponent $\theta$ comparable to ours \citep{Karimi2017} (and much larger than the one present in the absence of inertia in these materials). It has been proposed that depending on the amount of damping, different universality classes could exist \cite{Nicolas2016,Salerno2013}, but that for strongly underdamped systems the transition appears to become first order \citep{Karimi2017}. If confirmed, we speculate that the cause of the difference between amorphous solids and frictional interfaces is that avalanches are compact objects (having a fractal dimension $d_f > 1$) only in the latter case. If that would not be true, our assumption that the inertial noise within an avalanche is comparable to that occurring in a macroscopic slip event may not hold, possibly leading to different physics.

\paragraph{Acknowledgement}
\scriptsize

T.G.~was partly financially supported by The Netherlands Organisation for Scientific Research (NWO) by a NWO Rubicon grant number 680-50-1520. M.W.~thanks the Swiss National Science Foundation for support under Grant No.~200021-165509 and the Simons Foundation Grant ($\#$454953 Matthieu Wyart). We acknowledge an anonymous referee for useful comments on experimental validation.

\paragraph{Methods}
\scriptsize

We consider two ensembles, each consisting of independent realisations comprising an approximately square box characterised by $N$ blocks along the weak layer, with $N = 3^6$ ($300$ realisations) and $N = 3^6 \times 2$ ($1000$ realisations). The mechanical response is approximately incompressible, which allows us to focus on the shear response. The box is assumed periodic in horizontal direction. Quasi-static shear is applied by fixing the displacement of the bottom boundary to zero, while incrementing the displacement of the top boundary in very small steps (though we efficiently skip periods in which no yielding takes place, by homogeneously distributing the shear strain). Loading is stopped when the local strain exceeds a maximum.

After each step the energy is minimised according to the following equation of motion:
\begin{equation}
\label{eq:model:motion}
  \rho \; \vec{a}(\vec{r}) = \vec{\nabla} \cdot \bm{\sigma}\big(\bm{\varepsilon}(\vec{r})\big) - \alpha \vec{v}(\vec{r})
\end{equation}
(see Appendix~\ref{sec:nomenclature} for nomenclature). From left to right, this equation comprises (i) an inertial term, in which $\rho$ is the mass density and $\vec{a} = \partial_t^2 \vec{u}$ is the acceleration (where $\vec{u}$ is displacement and $t$ is time); (ii) the divergence of the stress tensor $\bm{\sigma}$; and (iii) a non-Rayleigh damping term, where $\alpha$ is the damping coefficient and $\vec{v} = \partial_t \vec{u}$ is the velocity. The stress $\bm{\sigma}$ follows from strain $\bm{\varepsilon}$ (which is the symmetric gradient of the displacement $\vec{u}$) using the constitutive model outlined in the main text. We set $\alpha$ such that kinetic energy is effectively leaked at the (periodic) boundaries.

Eq.~\eqref{eq:model:motion} is solved in the weak form by discretising in space and time. In space, we discretise using finite elements. These elements coincide with the elasto-plastic blocks along the weak layer, while the elastic domain is discretised using elements that are conveniently chosen to increase in size with increasing distance to the weak layer to save computational costs. We discretise in time using the velocity-Verlet protocol.

Note that we formulate our model under the small strain assumption. To respect this assumption but still acquire a decently long steady state response, we choose the yield strains to be very small. This fixes the absolute strain and stress values to be small, which we rescale for visualisation to be of order one. See Appendix~\ref{sec:model} for details. Furthermore, note that the numerical implementation is open-source \cite{GooseFEM,ElastoPlasticQPot} and that we have made all data underlying this manuscript freely available \cite{DeGeus2018_data}.

\normalsize

\bibliography{library}

\clearpage

\onecolumn
\clearpage
\justify
\appendix

\begin{center}
\Large\textbf{Appendix}
\end{center}

\section{Nomenclature}
\label{sec:nomenclature}

\def\arraystretch{1.5}
\begin{tabular}{lll}
$\vec{a}$ & vector & $\vec{a} = \sum_i a_i \vec{e}_i$ \\
$\bm{A}$ & second-order tensor & $\bm{A} = \sum_i \sum_j A_{ij} \vec{e}_i \vec{e}_j$ \\
$\bm{I}$ & second-order unit tensor & $\bm{I} = \sum_i \sum_j \delta_{ij} \vec{e}_i \vec{e}_j$ \\
$\bm{C} = \bm{A}^T$ & transpose of a second-order tensor & $C_{ij}= A_{ji} $ \\
$c = \mathrm{tr}\left(\bm{A}\right)$ & trace & $c = \sum_i A_{ii}$ \\
$c = \vec{a} \cdot \vec{b}$ & vector contraction (dot/inner product) & $c = \sum_i a_i b_i$ \\
$c = \bm{A} : \bm{B}$ & double tensor contraction (double dot/inner product) & $c = \sum_i \sum_j A_{ij} B_{ji}$ \\
$\bm{C} = \vec{\nabla} \vec{a}$ & gradient operator & $C_{ij} = \partial a_j / \partial x_i $ \\
$\vec{c} = \vec{\nabla} \cdot \bm{A}$ & divergence operator & $c_j = \partial A_{ij} / \partial x_i$ \\
$\underline{\vec{a}}$ & column of vectors & $\vec{a}_k$ \\
$\underline{\underline{A}}$ & matrix & $A_{kl}$ \\
\end{tabular}

\section{Model}
\label{sec:model}

\subsection{Constitutive model}
\label{si:constitutive}

\paragraph{Linear elasticity}

The constitutive model, as illustrated in Fig.~\ref{fig:model}, is based on linear elasticity. Such behaviour is provided by Hooke's law
\begin{equation}
  \bm{\sigma} \equiv \frac{K}{d} \mathrm{tr}\left( \bm{\varepsilon} \right) \bm{I} + G \bm{\varepsilon}_\mathrm{d}
  \label{si:constitutive:stress}
\end{equation}
where $K$ and $G$ respectively are the bulk and shear modulus, and $d$ is the number of dimensions ($d = 2$ in our case). Here $\bm{\varepsilon}$ is the (linear) strain tensor, defined as the symmetric gradient of the displacement $\vec{u}$, i.e.
\begin{equation}
  \bm{\varepsilon} \equiv \tfrac{1}{2} \left( \vec{\nabla} \vec{u} + ( \vec{\nabla} \vec{u} )^T \right)
\end{equation}
Finally, $\bm{\varepsilon}_\mathrm{d}$ is the strain deviator which contains all shear components of the strain, i.e.\ all strain components that do not lead to change of volume:
\begin{equation}
  \bm{\varepsilon}_\mathrm{d} \equiv \bm{\varepsilon} - \tfrac{1}{d} \, \mathrm{tr} ( \bm{\varepsilon} ) \, \bm{I}
\end{equation}
We assume that yielding is a shear transformation. The volumetric response (through $K$) is therefore assumed to be purely elastic. Plasticity is defined through the potential energy. To obtain the shear part of Eq.~\eqref{si:constitutive:stress} in an energetic picture, we need to introduce an equivalent shear strain
\begin{equation}
  \varepsilon \equiv \; \sqrt{ \tfrac{1}{2} \, \bm{\varepsilon}_\mathrm{d} : \bm{\varepsilon}_\mathrm{d} }
\end{equation}
This quantity thus characterises the magnitude of the shear strains encompassed in the strain deviator $\bm{\varepsilon}_\mathrm{d}$. With this, we retrieve Eq.~\eqref{si:constitutive:stress} with a quadratic potential energy for the deviatoric part of the strain:
\begin{equation}
  V(\varepsilon) = G \varepsilon^2
\end{equation}
(where $\bm{\sigma}_\mathrm{d} \equiv \partial V / \partial \bm{\varepsilon}_\mathrm{d}$). For completeness we introduce the work-conjugate equivalent shear stress $\sigma^2 \equiv 2 \bm{\sigma}_\mathrm{d} : \bm{\sigma}_\mathrm{d}$.

\paragraph{Elasto-plasticity}

Following \citet{Jagla2017}, a shear transformation is introduced by using a manifold of quadratic potentials, given by:
\begin{equation}
  V \Big(
    \varepsilon_\mathrm{y}^{(i)} \leq \varepsilon < \varepsilon_\mathrm{y}^{(i+1)}
  \Big)
  =
  G \, \bigg[\,
    \Big[\, \varepsilon - \varepsilon_\mathrm{min}^{(i)} \,\Big]^2
    -
    \Big[\, \Delta \varepsilon_\mathrm{y}^{(i)} \,\Big]^2
  \,\bigg]
\end{equation}
see Fig.~\ref{fig:model}. The elastic response is always governed by the shear modulus $G$. The intersections of the potentials are set by the sequence of yield strains $\varepsilon_\mathrm{y}^{(i)}$ (where $\varepsilon_\mathrm{y}^{(i+1)} > \varepsilon_\mathrm{y}^{(i)}$). Furthermore, $\Delta \varepsilon_\mathrm{y}^{(i)} \equiv (\varepsilon_\mathrm{y}^{(i+1)} - \varepsilon_\mathrm{y}^{(i)})/2$ and $\varepsilon_\mathrm{min}^{(i)} \equiv \varepsilon_\mathrm{y}^{(i)} + \Delta \varepsilon_\mathrm{y}^{(i)}$. Compared to Ref.~\citep{Jagla2017} our model remains isotropic, which corresponds to a yield strain that bounds the elastic domain using a sphere in principal deviatoric strain space. Finally, we obtain the following expression for the stress tensor:
\begin{equation}
  \bm{\sigma} ( \bm{\varepsilon} )
  =
  \frac{K}{d} \mathrm{tr}\left( \bm{\varepsilon} \right) \bm{I}
  +
  G \, \Big[\, \varepsilon - \varepsilon_\mathrm{min}^{(i)} \,\Big] \;
  \bm{N}_\mathrm{d}
  \qquad
  \mathrm{for}
  \;
  \varepsilon_\mathrm{y}^{(i)} \leq \varepsilon < \varepsilon_\mathrm{y}^{(i+1)}
\end{equation}
whereby the direction of shear is contained in
\begin{equation}
  \bm{N}_\mathrm{d} \equiv \frac{\bm{\varepsilon}_\mathrm{d}}{\varepsilon}
\end{equation}
Note finally that $\varepsilon_\mathrm{p} = \varepsilon_\mathrm{min}^{(i)}$ may be interpreted as a plastic strain, and thus that $\varepsilon_\mathrm{e} = \varepsilon - \varepsilon_\mathrm{min}^{(i)}$ corresponds to an elastic strain.

\paragraph{Parameters}

\begin{figure}
  \centering
  \includegraphics[width=.4\textwidth]{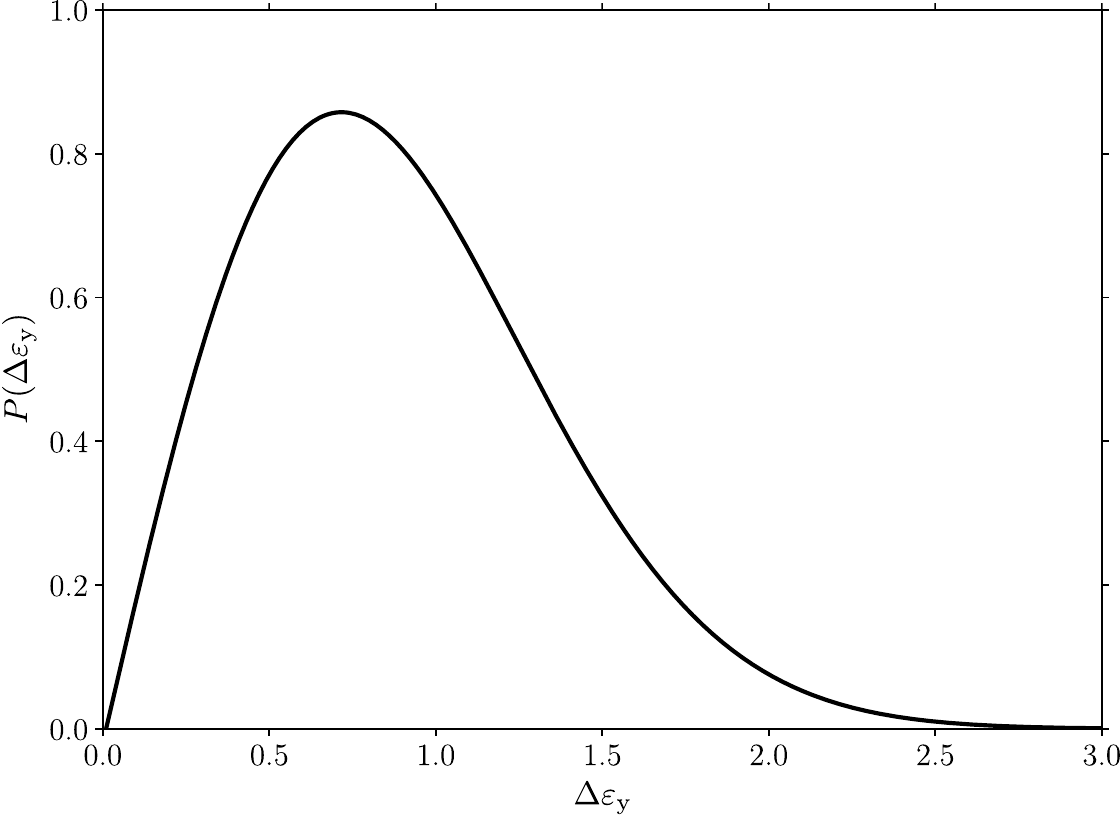}
  \caption{Distribution of yield strains $\Delta \varepsilon_\mathrm{y}$ as given by Eq.~\eqref{eq:yield-strains}. Practically the yield strains of a block are drawn for the entire strain history before starting the simulation. This mimics a surface roughness that does not evolve during frictional sliding. It also allows easy post-processing as the simulations rely on random input, but themselves are not stochastic.}
  \label{fig:si:yield-strains}
\end{figure}

The yield strains are drawn from a Weibull distribution
\begin{equation}
  P(\Delta\varepsilon_\mathrm{y}) = k \; (\Delta\varepsilon_\mathrm{y})^{k-1} \exp\left[ -(\Delta\varepsilon_\mathrm{y})^k \right]
  \label{eq:yield-strains}
\end{equation}
for which we use $k = 2$, see Fig.~\ref{fig:si:yield-strains}. To avoid difficulties in our algorithm (which is freely available \citep{ElastoPlasticQPot}) we add a small offset $\delta_\mathrm{y}$ to each drawn yield strain, such that $P(\Delta\varepsilon_\mathrm{y} < \delta_\mathrm{y}) = 0$. To minimise the duration (in strain) of the transient initial loading preceding the steady state (highlighted in Fig.~\ref{fig:typical}(a)), the first yield strain of each block is taken from a uniform distribution $\Delta\varepsilon_\mathrm{y} = [\delta_\mathrm{y},1)$. Incompressibility is approximated by $K/G = 10$, whereby the shear modulus $G = 1$. Furthermore, the shear wave speed $c_s = \sqrt{G/(2\rho)} = 1/\sqrt{2}$ (note that the factor $1/\sqrt{2}$ appears because of our definition of the shear modulus). A crucial final point is that the absolute strains (and stresses) are fixed by the yield strains. To stay, as much as possible, within the small strain limit we scale the yield strains by $\varepsilon_0 = 10^{-3} / 2$ (and use $\delta_\mathrm{y} = 10^{-5} / 2$). All strains and stresses that appear in diagrams have been rescaled by this typical strain and corresponding typical stress making them $\mathcal{O}(1)$.

\subsection{Finite element discretisation}
\label{si:fem}

\paragraph{Overview}

We solve Eq.~\eqref{eq:model:motion} by discretising space according to the Finite Element Method (FEM). Besides our elasto-plastic blocks, we discretise also the elastic region using quadrilateral elements. This is illustrated in Fig.~\ref{fig:model}, where it is observed that we systematically coarsen the regions that present less interesting physics to reduce computational costs. FEM treats Eq.~\eqref{eq:model:motion} in a weak sense. The resulting volume integral is evaluated element-by-element using numerical quadrature, in our case using four Gauss points. It is at these points that the stress and strain are evaluated, and thus where the potentials that are illustrated in Fig.~\ref{fig:model} are defined. To fix the disorder to the scale of the blocks, all four Gauss points in one element get the same local potential energy landscape. The discretised weak form of Eq.~\eqref{eq:model:motion} is, furthermore, discretised in time using the velocity-Verlet protocol. This results in incremental updates for the nodal velocity $\underline{\vec{v}}$ and displacement $\underline{\vec{u}}$, based the nodal acceleration $\underline{\vec{a}}$ that results from solving the discrete equation of motion taking into account the fixed displacement and periodic boundary conditions (as illustrated in Fig.~\ref{fig:model}). Note that our FEM code is also freely available \citep{GooseFEM}.

\paragraph{Weak form}

The crux of the Finite Element Method is to solve Eq.~\eqref{eq:model:motion} in its weak form. For this, one has to satisfy
\begin{equation}
  \int\limits_\Omega
    \rho(\vec{r})\; \delta\vec{u}(\vec{r}) \cdot \vec{a}(\vec{r}) \;
  \mathrm{d}\Omega
  =
  \int\limits_\Omega
    \delta\vec{u}(\vec{r})
    \cdot
    \Big[\,
      \vec{\nabla}
      \cdot
      \bm{\sigma}(\vec{r})
      -
      \alpha \vec{v}(\vec{r})
    \,\Big] \;
  \mathrm{d}\Omega
\end{equation}
for any possible test function $\delta\vec{u}(\vec{r})$. Note that $\Omega$ is the volume of the box. We note once more that the stress $\bm{\sigma}(\vec{r})$ depends in some specific way on the strain $\bm{\varepsilon}(\vec{r})$: the symmetric gradient of the displacement $\vec{u}(\vec{r})$ (see Appendix~\ref{si:constitutive} for details). In order to be able to evaluate the integral element-by-element, partial integration is employed next. This results in
\begin{equation}
\label{eq:fem:weak}
  \int\limits_\Omega
    \rho(\vec{r})\; \delta\vec{u}(\vec{r}) \cdot \vec{a}(\vec{r}) \;
  \mathrm{d}\Omega
  =
  -
  \int\limits_\Omega
    \big[\, \vec{\nabla} \delta\vec{u}(\vec{r}) \,\big]
    :
    \bm{\sigma}(\vec{r}) \;
  \mathrm{d}\Omega
  -
  \int\limits_\Omega
    \alpha\; \delta\vec{u}(\vec{r}) \cdot \vec{v}(\vec{r}) \;
  \mathrm{d}\Omega
  \qquad
  \forall \; \delta\vec{u}(\vec{r}) \in \mathbb{R}^d
\end{equation}
whereby the boundary integral, that incorporates the external forces that appear from partial integration has been omitted as its contribution is irrelevant because we fix all displacements along the top and bottom boundaries, and assume periodicity along the rest of the boundary. The discretised external forces needed to sustain the prescribed displacements can be easily retrieved, as discussed in Appendix~\ref{si:event}.

\paragraph{Discretisation in space}

\begin{figure}[htp]
  \centering
  \includegraphics[width=.6\textwidth]{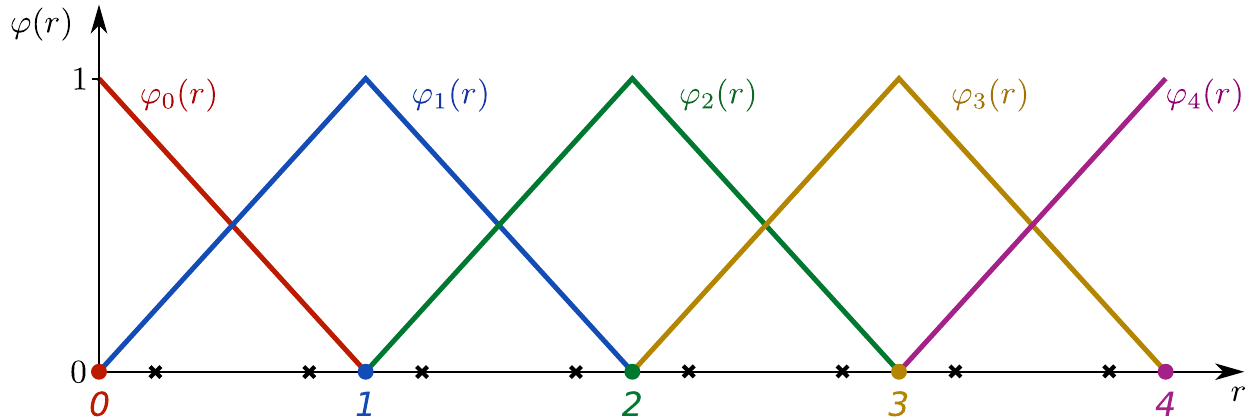}
  \caption{Shape functions in one dimension. The shape functions belonging to each node are shown using a different colour, and are only shown where they are non-zero. The nodes are shown using markers, and the integration points using crosses. In one dimension each element is bounded by two nodes.}
  \label{fig:shape-functions-1d}
\end{figure}

The problem is now discretised in space using a set of nodes that are connected through elements. Shape functions $\varphi_k(\vec{r})$ are used to interpolate the nodal displacement and test functions throughout the discretised domain $\Omega^h$. Note thereby that $\varphi_k(\vec{r})$ is locally supported, being non-zero only in the elements that contain the node $k$ (see Fig.~\ref{fig:shape-functions-1d} for a one-dimensional example). Furthermore, the shape functions constitute to a partition of unity. Their expression is standard, and for our four-noded quadrilateral elements they are bilinear.

For the displacement field and test functions, we thus have that
\begin{alignat}{2}
  \vec{u}(\vec{r})
  &\approx
  \vec{u}^h(\vec{r})
  &&=
  \sum_{k} \varphi_k (\vec{r}) \; \vec{u}_k
  \label{eq:fem:shape-u}
  \\
  \delta\vec{u}(\vec{r})
  &\approx
  \delta\vec{u}^h(\vec{r})
  &&=
  \sum_{k} \varphi_k (\vec{r}) \; \delta\vec{u}_k
  \label{eq:fem:shape-deltau}
\end{alignat}
where $k$ loops over all nodes. When applied to Eq.~\eqref{eq:fem:weak} we get
\begin{equation}
  \underbrace{
    \int\limits_{\Omega^h}
      \rho(\vec{r})\; \varphi_k(\vec{r})\; \varphi_l(\vec{r}) \;
    \mathrm{d}\Omega^h
  }_{\displaystyle
    M_{kl}(\vec{r})
  } \;
  \vec{a}_l
  =
  -
  \underbrace{
    \int\limits_{\Omega^h}
      \big[\, \vec{\nabla} \varphi_k(\vec{r}) \,\big]
      :
      \bm{\sigma}(\vec{r}) \;
    \mathrm{d}\Omega^h
  }_{\displaystyle
    \vec{f}_k(\vec{r})
  }
  -
  \underbrace{
    \int\limits_{\Omega^h}
      \alpha(\vec{r})\; \varphi_k(\vec{r})\; \varphi_l(\vec{r}) \;
    \mathrm{d}\Omega^h
  }_{\displaystyle
    D_{kl}(\vec{r})
  } \;
  \vec{v}_l
  \label{eq:dynamics:system}
\end{equation}
Which is automatically satisfied for all nodal test functions. The integrals are finally evaluated at a discrete set of quadrature points. For our quadrilateral elements this step is exact when using four so-called Gauss points. We make one approximation here for the integrals that result in $M_{kl}$ and $D_{kl}$, by choosing equally weight quadrature points that coincide with the nodes. By making this approximation both matrices become diagonal, and their numerical treatment, including inversion, very cheap. Physically this corresponds to concentrating the mass as point masses at the nodes (whereby the mass still depends on the corresponding volume). We finally note that the strains at the Gauss points, that are needed to compute the stresses there, are obtained from the interpolation of the nodal displacements using Eq.~\eqref{eq:fem:shape-u}. Finally, we introduce a short hand notation that reads
\begin{equation}
  \underline{\underline{M}}\, \underline{\vec{a}} = - \underline{\vec{f}}_\mathrm{int} = - \underline{\vec{f}} - \underline{\underline{D}}\, \underline{\vec{v}}
  \label{eq:fem:system}
\end{equation}

\paragraph{Discretisation in time}

To solve the second order differential equation in time, we proceed by discretising time. For this we employ the velocity-Verlet protocol, which:
\begin{enumerate}
  \item Computes the position at time $t^{(n+1)} = t^{(n)} + \Delta_t$:
  \begin{equation}
    \underline{\vec{u}}^{(n+1)}
    =
    \underline{\vec{u}}^{(n)} + \Delta_t \underline{\vec{v}}^{(n)} + \tfrac{1}{2} \Delta_t^2 \underline{\vec{a}}^{(n)}
  \end{equation}
  \item Estimates the velocity at time $t^{(n+1)} = t^{(n)} + \Delta_t$ (by solving Eq.~\eqref{eq:fem:system}):
  \begin{equation}
    \hat{\underline{\vec{v}}}^{(n+1)}
    =
    \underline{\vec{v}}^{(n)}
    +
    \tfrac{1}{2} \Delta_t \Big[\,
      \underline{\vec{a}}^{(n)} + \underline{\vec{a}} \big( \underline{\vec{u}}^{(n+1)} , \underline{\vec{v}}^{(n)} + \Delta_t \underline{\vec{a}}^{(n)} , t^{(n+1)} \big) \,
    \Big]
  \end{equation}
  \item Corrects $\hat{\underline{\vec{v}}}^{(n+1)}$ (by solving Eq.~\eqref{eq:fem:system}):
  \begin{equation}
    \underline{\vec{v}}^{(n+1)}
    =
    \underline{\vec{v}}^{(n)}
    +
    \tfrac{1}{2} \Delta_t \Big[\,
      \underline{\vec{a}}^{(n)} + \vec{a} \big( \underline{\vec{u}}^{(n+1)} , \underline{\hat{\vec{v}}}^{(n+1)} , t^{(n+1)} \big) \,
    \Big]
  \end{equation}
  \item Computes $\underline{\vec{a}}^{(n+1)}$ by solving Eq.~\eqref{eq:fem:system} (using $\underline{\vec{u}}^{(n+1)}$ and $\underline{\vec{v}}^{(n+1)}$).
\end{enumerate}

\paragraph{Parameters}

To set the background damping such that waves whose wavelength is longer than the system's size are critically damped we use the following one-dimensional wave equation
\begin{equation}
  \rho \partial_t^2 u = \tfrac{1}{2} G \partial_x^2 u - \alpha \partial_t u
  \label{eq:si:wave}
\end{equation}
Critical damping is found for wave numbers
\begin{equation}
  q = \alpha / ( 2 c_s \rho)
\end{equation}
substituting $q = 2 \pi / ( N h )$ gives us the value of $\alpha$ that we seek.

The time step is taken much smaller than the time needed for a shear wave to travel to the shortest length scale in our problem: the block's size. Consequently we take $\Delta_t = (1/(\sqrt{2} c_s q_h)) / 10$, with where $q_h = 2 \pi / h$.

\subsection{Event-driven protocol}
\label{si:event}

The box (e.g.\ in Fig.~\ref{fig:model}) is sheared, in simple shear, by prescribing the displacement of the top boundary. We stop the simulation when the local equivalent shear strain reaches $\varepsilon = 0.5$ anywhere in the system (note that this is the absolute strain, before rescaling by the typical strain $\varepsilon_0$). As we seek to carefully measure the avalanches triggered by a local yielding event, we prescribe a very small equivalent shear strain change ($\delta_\varepsilon = 10^{-7}$, also in terms of the absolute strain) during each loading increment. This coincides with the quasi-static protocol. The protocol is thereby that we affinely displace the entire box, such that the strain increment is homogeneous (see below). The top (and bottom) boundaries are then kept fixed, while the rest of the system evolves until energy has been minimised.

To enhance efficiency we make use of the relatively large strain intervals in which the entire box responds elastically. Since we drive very slowly and we know exactly how to distribute the strains to reach equilibrium (simply homogeneously in this case). We can thus transverse the entire elastic regime in one step, allowing us to run an event-driven quasi-static loading protocol. This protocol consists of two steps. In the first step, an affine displacement is added to the entire box such that the point that was closest to yield, is brought to the verge of yielding. In particular, the affine displacement is applied such that the increment in equivalent shear strain, $\Delta \varepsilon$, satisfies:
\begin{equation}
  \min\limits_{\vec{r}} \left[
    \varepsilon_\mathrm{y} (\vec{r}) - \big( \varepsilon^{(n+1)} (\vec{r}) + \Delta \varepsilon \big)
  \right] = \delta_\varepsilon / 2
  \label{eq:si:event:min}
\end{equation}
(below we describe how this translates to an affine displacement increment). Since this step is purely elastic (and the displacement is affine), energy is instantaneously minimised. Then, in the second step, a small `kick' $\Delta\varepsilon = \delta_\varepsilon$ is given to the system (again by applying an affine displacement), that causes yielding in at least one point.

An affine shear displacement field $\Delta u_x (y) = y \, \Delta \gamma$ is applied (where $x$ is the horizontal coordinate and $y$ is the vertical coordinate). This leads to the following local strain deviator:
\begin{equation}
  \bm{\varepsilon}^{(n+1)}_\mathrm{d} (\vec{r})
  =
  \bm{\varepsilon}^{(n)}_\mathrm{d} (\vec{r})
  +
  \begin{bmatrix}
    0 &
    \Delta \gamma \\
    \Delta \gamma &
    0
  \end{bmatrix}
  =
  \begin{bmatrix}
     \varepsilon_\mathrm{ss}^{(n)}(\vec{r}) &
     \varepsilon_\mathrm{ps}^{(n)}(\vec{r}) + \Delta \gamma \\
     \varepsilon_\mathrm{ps}^{(n)}(\vec{r}) + \Delta \gamma &
    -\varepsilon_\mathrm{ss}^{(n)}(\vec{r})
  \end{bmatrix}
\end{equation}
where $\varepsilon_\mathrm{ss}$ and $\varepsilon_\mathrm{ps}$ indicate the local contributions in simple shear (ss) and pure shear (ps) -- the two principle deviatoric strains. This gives us an expression for the equivalent shear strain
\begin{equation}
  \left( \varepsilon^{(n+1)}(\vec{r}) \right)^2
  =
  \left( \varepsilon^{(n)}(\vec{r}) + \Delta \varepsilon \right)^2
  =
  \left( \varepsilon_\mathrm{ss}^{(n)}(\vec{r}) \right)^2
  +
  \left( \varepsilon_\mathrm{ps}^{(n)}(\vec{r}) + \Delta \gamma \right)^2
\end{equation}
which can be solved exactly for $\Delta \gamma$, thereby taking the strain components from the point that is closest to yielding (see Eq.~\eqref{eq:si:event:min}).

The kick in strain is followed by an energy minimisation, using the equation of motion in Eq.~\eqref{eq:fem:system} until all residual forces are sufficiently small. In particular we satisfy
\begin{equation}
  \frac{\sum_k \left| \vec{F}_k - \vec{f}^\mathrm{int}_k \right|}{\sum_k \left| \vec{F}_k \right|} \leq 10^{-5}
\end{equation}
where the reaction forces, $\vec{F}_k$, measured at the nodes whose displacement is fixed (i.e.\ those at the top and bottom boundaries), are used for normalisation. They are
\begin{equation}
  \vec{F}_k =
  \begin{cases}
    \vec{f}^\mathrm{int}_k \qquad& \text{if}\; k \in \text{fixed displacement boundary} \\
    \vec{0}   \qquad& \text{otherwise}
  \end{cases}
\end{equation}

\subsection{Triggering}
\label{si:trigger}

We measure the response at different stresses above $\sigma_c$ by manually triggering events after macroscopic slip (during which all blocks yielded at least once). In particular, we trigger at different $\sigma$ above $\sigma_c$ after macroscopic slip, as is illustrated in Fig.~\ref{fig:typical}(a,b). This state can be reached exactly by applying an affine deformation to the last equilibrium state following each macroscopic slip (and an arbitrary number of avalanches) for which the stress is still smaller than $\sigma$, as sketched in Fig.~\ref{fig:typical}(b). This protocol provides us with one equilibrium state following each macroscopic slip (provided that it ended at a sufficiently low stress, and that the next macroscopic slip nucleated at a sufficiently high stress). To acquire statistics we obtain more than one measurement from each relevant equilibrium state by triggering different blocks per realisation: each $73^\mathrm{th}$ block along the weak layer.

\begin{figure}[htp]
  \centering
  \includegraphics[width=.3\textwidth]{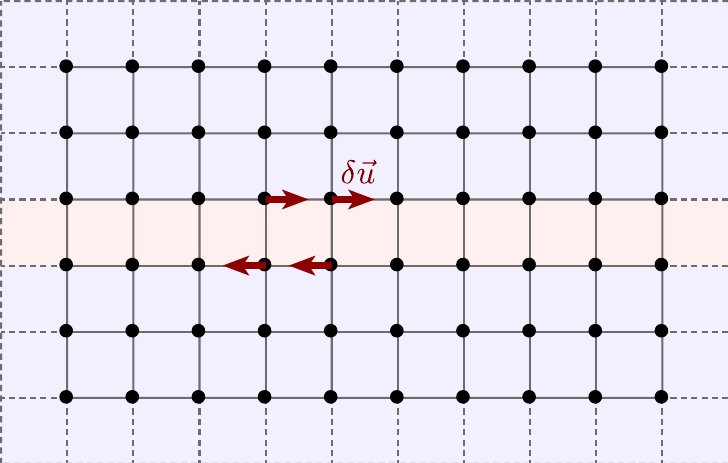}
  \caption{Sketch of the triggering protocol. A shear displacement fluctuation is applied to a randomly selected block along the weak layer at one time step $t = t_t$; this fluctuation is temporary, only lasting one time step, the boundary conditions are not changed in any way. If the configuration were elastic, the system would simply snap back to the original configuration. In the elasto-plastic configuration, however, an avalanche of yielding events can be triggered, leading to a new equilibrium configuration (corresponding to a lower energy state) obeying the same boundary conditions.}
  \label{fig:trigger}
\end{figure}

We trigger the event by temporarily applying a displacement fluctuation to the selected block, see Fig.~\ref{fig:trigger}. Note that the boundary conditions are not changed in any way, the system is free to possibly reach a new minimum governed by the same boundary conditions. The amplitude of the displacement fluctuation is such that the strain in the block is just above the yield strain; the same strain `kick' is used as in Appendix~\ref{si:event}.

\section{(Typical) Response}
\label{sec:response}

\subsection{A single plastic event}
\label{si:single_event}

When a single block yields, it releases its built-up potential energy, accompanied by an increase in shear strain. This triggers dipolar force field on the surrounding blocks, which leads a change of stress whose amplitude decays in space as $1/r^{d}$, with $r$ being the distance to the yielding block and $d$ the number of dimensions ($d = 2$ in our case). Along different directions this change is positive or negative, but along our weak layer the stress is strictly increased. This corresponds to the classical results by \citet{Eshelby1956}, accessibly summarised in a recent review of elasticity \citep{Clouet2018}.

Inertia introduces a second, transient, effect. Upon the sudden release of elastic energy, the surrounding blocks transiently experience the effect of a series of monopolar forces. This causes a series of temporary stress overshoots, followed by temporary stress undershoots in the surrounding blocks, whereby the amplitude of the stress overshoot decays in space as $1/r^{d-1}$.

We can observe these classical results in our system by triggering yielding of a single block embedded in an otherwise homogeneous elastic system (using the same simple-shear drive as is being used in the main text, see Fig.~\ref{fig:model}). Fig.~\ref{fig:single_event} shows the response in the neighbouring blocks along the horizontal direction. The scaling of the permanent stress increase measured at long time (denoted by $\Delta \sigma$, in blue) and the temporary maximum stress overshoot (denoted by $\sigma_I$, in red) are consistent with our prediction.

\subsection{Ballistic rupture front}
\label{sec:time-evolution}

\begin{figure}[htp]
  \begin{minipage}[b]{.49\textwidth}
  \centering
  \includegraphics[width=\textwidth]{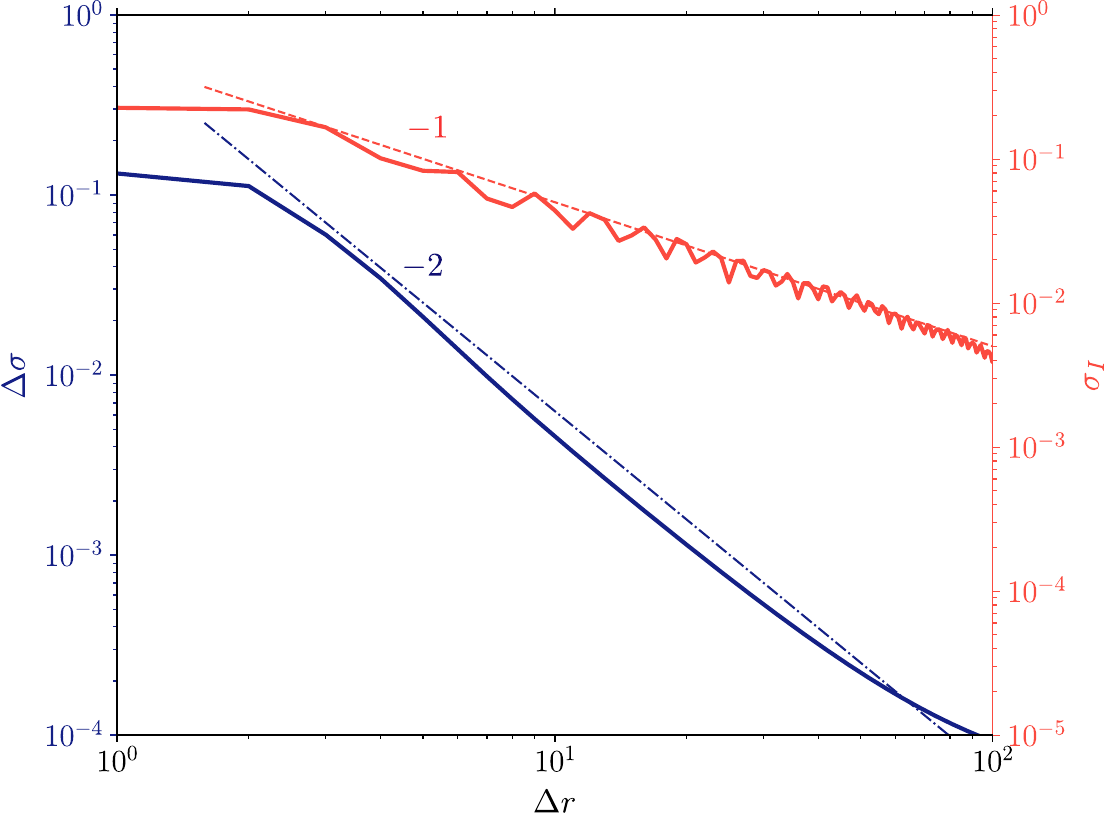}
  \captionof{figure}{The effect of yielding of a single block in an otherwise homogeneous elastic box for which $N = 3^6$, in terms of: the permanent stress redistribution (defined as the difference between the stress at time $t = \infty$ and the yield stress, in blue), and the temporary stress overshoot (defined as the difference between maximum stress at any time and the yield stress, in red); both as a function of horizontal distance to the yielding block (in number of blocks). Note that for the single elastic block, $\Delta \varepsilon_\mathrm{y}^{(i)}$ is taken from a delta distribution. The dashed lines mark the power law scaling with the indicated exponents.}
  \label{fig:single_event}
  \end{minipage}
  \hfill
  \begin{minipage}[b]{.49\textwidth}
  \centering
  \includegraphics[width=\textwidth]{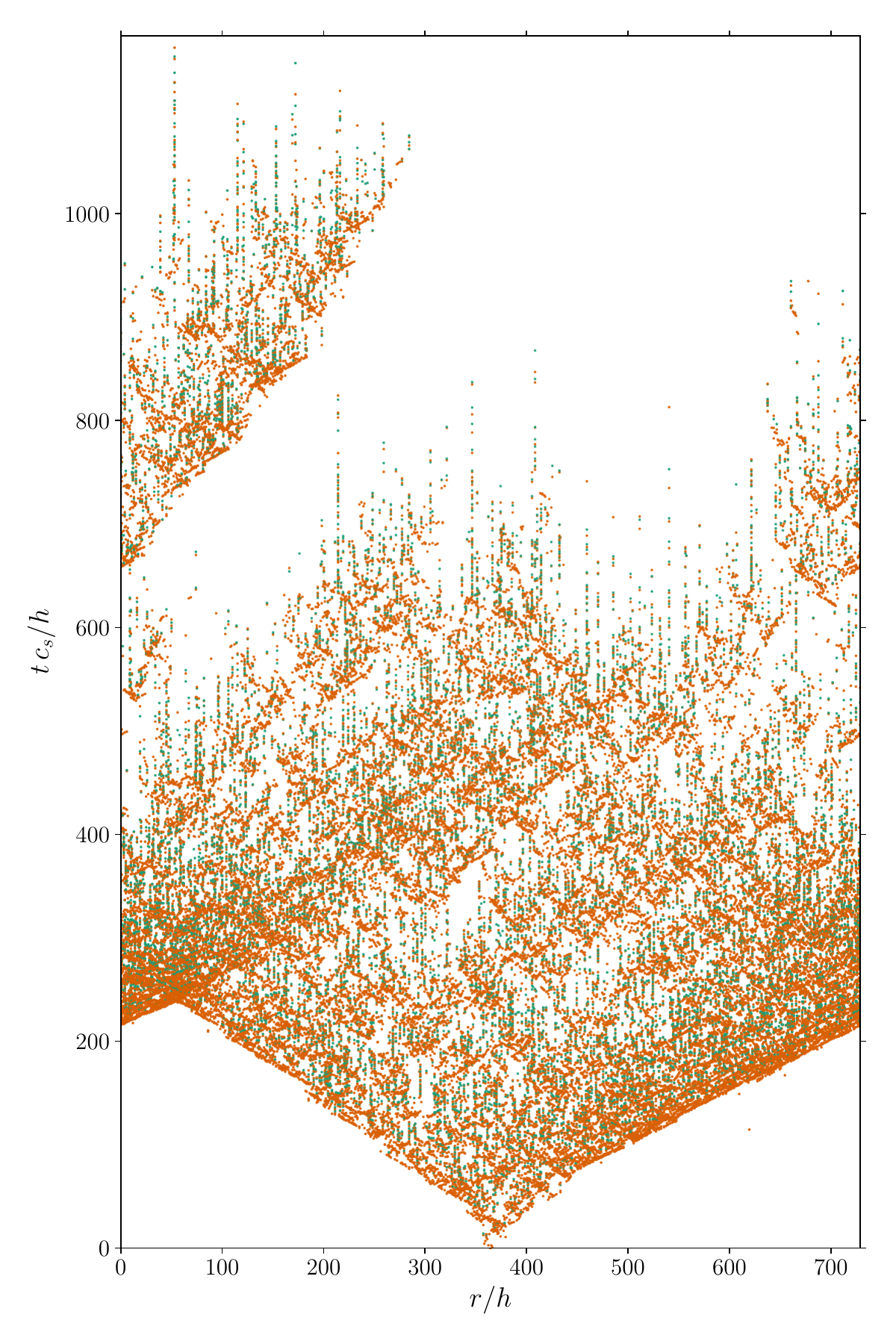}
   \captionof{figure}{Yield-map of the time evolution of a single system-spanning-avalanches that started approximately at $\sigma_n$ during the (regular) event-driven protocol. The horizontal axis corresponds to the position along the weak layer, normalised by the linear size $h$ of each block. The vertical axis corresponds to the time, normalised by the time it takes a shear wave to travel the distance of one block ($c_s$ is the shear wave speed, see Appendix~\ref{si:constitutive}). Each marker corresponds to a yielding event (orange forward, green backward). Note that to ease interpretation, the position is shifted, by making use of the periodicity, such that the block that yielded first is centred. Also note that $N = 3^6$.}
  \label{fig:time-evolution}
  \end{minipage}
\end{figure}

To support our picture of an avalanche that nucleates a fracture, it is instructive to consider the time evolution during a single system-spanning event (corresponding to macroscopic slip characterised by a macroscopic stress drop, see Fig.~\ref{fig:typical}(a)). In particular we expect to first see the avalanche as a fractal object (that is rather compact as $d_f > 1$). Beyond a critical radius $A_c$ this object transitions into a clear rupture front. Since the system is finite, $N = 3^6$ for the results below, nucleation happens at a relatively high stress, or a small nucleation radius $A_c$. We now consider a typical time evolution, in Fig.~\ref{fig:time-evolution}, whereby each marker corresponds to a local yield event. Indeed, after the first yielding event, an avalanche is observed in which the blocks yield over and over. Then, after some time, the avalanche succeeds in nucleating a fracture that is characterised by a well-defined rupture front. Although not studied here, we remark that the rupture front is supersonic, its velocity $\simeq 2 c_s$ (which is, in accordance with the laws of elasticity, below the compressive wave speed, that for the used elastic constants is $\sqrt{10}\, c_s$, see Appendix~\ref{si:constitutive}). After this front has crossed the entire box, yielding continues, corresponding to macroscopic slip.

\section{Stability \& number of avalanches}
\label{sec:stability}

\subsection{Argument for pseudo-gap \texorpdfstring{$P(x_\sigma) \sim (x_\sigma)^\theta$}{P(x)}}
\label{si:stability}

The probability that a block will become unstable by a stress increase $\Delta_\sigma$ is
\begin{equation}
  P( x_\sigma < \Delta_\sigma ) \sim \int\limits_0^{\Delta_\sigma} (x_\sigma)^\theta \;\mathrm{d} x_\sigma \sim (\Delta_\sigma)^{\theta + 1}
\end{equation}
The stress increase caused by the failure of a single block decays in space as
\begin{equation}
  \Delta_\sigma \sim 1 / r^q
\end{equation}
with $q = d-1$ for the inertial stress overshoot and $q = d$ for the permanent stress increase, see Fig.~\ref{fig:single_event}. The number of blocks that will be destabilised by the failure of a block therefore reads
\begin{equation}
  n_f \sim
  \int\limits_h^R r^{-q(\theta + 1) + d - 2} \;\mathrm{d}r \sim
  \begin{cases}
    \ln r \Big|_h^R &\mathrm{for}\; d - q(\theta +1) = 1 \vspace*{.5em}\\
    r^{-q(\theta+1) + d-1} \Big|_h^R &\mathrm{otherwise}
  \end{cases}
\end{equation}
The only way that $n_f$ does not diverge for $R \rightarrow \infty$, for neither the inertial stress overshoot with $q = d-1$ nor the permanent stress increase with $q = d$, is when $\theta > 0$. Note that the number of failing blocks remains finite because the microscopic length scale, $h$, is finite. This proves that $P(x_\sigma)$ displays a pseudo-gap, with $\theta > 0$, or a gap (though the scenario of a gap is excluded by our data).

\subsection{Distance to yielding}
\label{sec:theta}

\begin{figure}[htp]
  \centering
  \includegraphics[width=.8\textwidth]{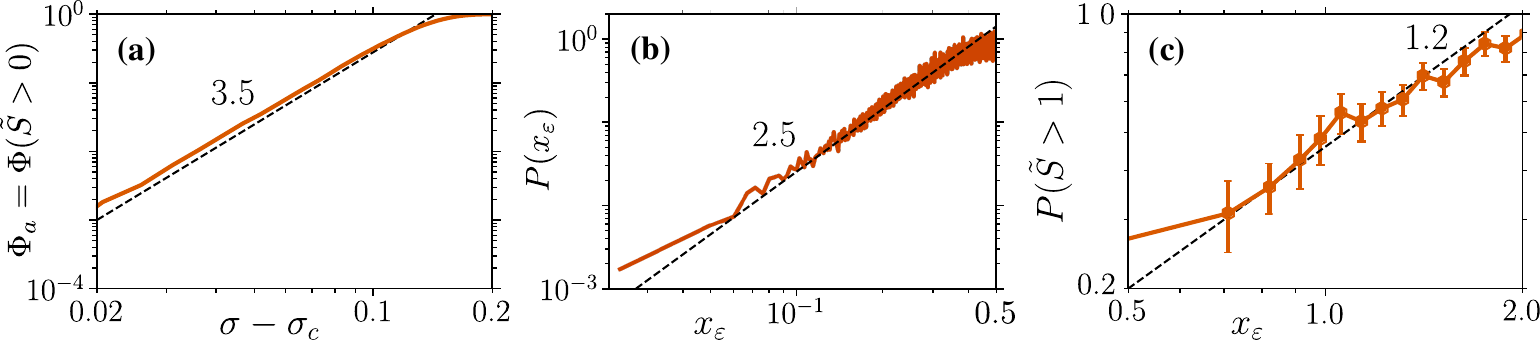}
  \caption{\textbf{(a)} Cumulative density of events as a function of the stress increase $\Delta_\sigma = \sigma - \sigma_c$ (without biasing for if such an event triggers an avalanche). \textbf{(b)} Distribution of the local stress increase needed to trigger yielding, at $\sigma = \sigma_c$. \textbf{(c)} Probability of triggering an avalanche as a function of $x_\varepsilon$. The combined distribution corresponds to the probability to trigger an avalanche upon a certain stress increase $x_\varepsilon$: $\theta \simeq 2.5 + 1.2 = 3.7$ (cf.\ Fig.~\ref{fig:theta}). The dashed line marks, in each figure, the power law scaling with the indicated exponent.}
  \label{fig:stability}
\end{figure}

We distinguish: (I) the distance to yielding of individual blocks from (II) the distance to triggering an avalanche. Namely, blocks can yield without triggering an avalanche, which can alter the relevant value of exponent $\theta$.  We measure both distributions independently. For the first distribution (I) we measure, for each block $i$, the additional amount of stress needed to yield: $x_{\varepsilon}^{(i)} \equiv \varepsilon_\mathrm{y}^{(i)} - \varepsilon^{(i)}$ (without loss of generality we directly use strain, which is uniquely related to the stress using the shear modulus: $x_\sigma \sim x_\varepsilon$). The distribution displays a pseudo-gap
\begin{equation}
  P(x_\varepsilon) \sim (x_\varepsilon)^{\theta^\prime}
\end{equation}
with an exponent $\theta^\prime \simeq 2.5$, as shown in Fig.~\ref{fig:stability}(b) and~\ref{fig:si:rho(x):correlate}(a,b). Note that this measurement is consistent with the scaling of the cumulative probability of yield events upon a stress increase $\Delta_\sigma = \sigma - \sigma_c$, that scales as $(\Delta_\sigma)^{\theta^\prime + 1}$ as shown in Fig.~\ref{fig:stability}(a).

For the second distribution (II), we measure the probability that an avalanche occurs (that yielding occurs more than once) when manually triggering yielding of a block at a certain $x_\varepsilon$. We find that this probability scales like $(x_\varepsilon)^{1.2}$, see Fig.~\ref{fig:stability}(c). These measurements are thus consistent with $\theta \simeq 2.5 + 1.2 = 3.7$ (cf.\ Fig.~\ref{fig:theta}).

\subsection{Microscopic details matter}
\label{si:distro2}

To test if the value of $\theta$ is universal we inspect the energy landscape around the configuration that is in mechanical equilibrium after macroscopic slip. In our case, this local energy landscape is fully defined by the yield strains in each block. On average, we find that the yield strains are large (40\% larger than the average yield strain drawn from the yield strain distribution, which we set to $1$ after rescaling the data, see Appendix~\ref{si:constitutive}). Since $\theta$ characterises the blocks close to yielding, we focus on the energy landscape around the blocks displaying a small $x_\varepsilon^{(i)}$. To this end, we compute the average yield strain bounding the local energy minimum of those blocks characterised by a small $x_\varepsilon^{(i)}$. We also compute the average yield strains bounding the next and previous local minima in the same block, as well as its left and right neighbours. Following \citep{DeGeus2015a}, we can compute:
\begin{equation}
  \big\langle \Delta\varepsilon_\mathrm{y}^{(\Delta i)} (\Delta \vec{r}) \big\rangle
  =
  \left(
    \sum\limits_{\vec{r}}\sum\limits_{i}
    \tilde{x}_\varepsilon^{(i)}(\vec{r})
    \;\;
    \Delta\varepsilon_\mathrm{y}^{(i+\Delta i)} (\vec{r} + \Delta \vec{r})
  \right)
  /
  \left(
    \sum\limits_{\vec{r}}\sum\limits_{i}
    \tilde{x}_\varepsilon^{(i)}(\vec{r})
  \right)
\end{equation}
To bias this weighted average to those blocks that are closest to yield, we use
\begin{equation}
  \tilde{x}_\varepsilon = \lfloor 0.2 - x_\varepsilon \rfloor
\end{equation}
where
\begin{equation}
  \lfloor \bullet \rfloor \, = \tfrac{1}{2} \big( \, \bullet + |\bullet| \, \big)
\end{equation}
takes only the positive part of $\bullet$. As a reference, we include the average yield strain without biasing for small distance to yielding ($\tilde{x}_\varepsilon = x_\varepsilon$, denoted without any subscript). All results are normalised by the ensemble average yield strain $\langle \Delta \varepsilon_\mathrm{y} \rangle_0 \approx 1$.

\begin{figure}
  \centering
  \includegraphics[width=.9\textwidth]{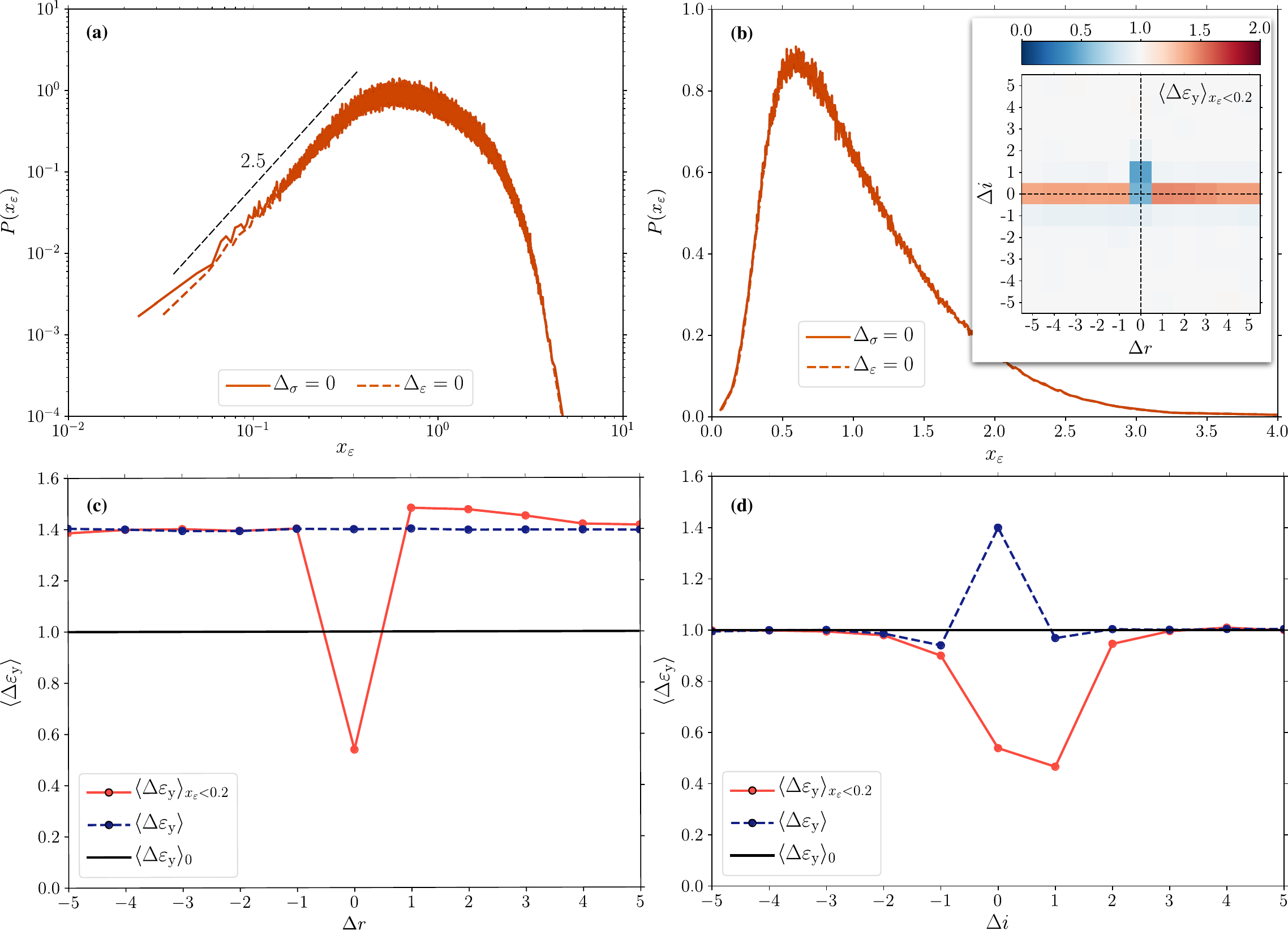}
  \caption{Distribution of the local distance to yielding, $P(x_\varepsilon)$ (see text for definition), after a macroscopic slip (directly after a macroscopic slip, indicated by $\Delta_\varepsilon = 0$, and at $\Delta_\sigma = 0$) on \textbf{(a)} a logarithmic and \textbf{(b)} a linear scale. The dashed line in (a) marks the power law scaling with the indicated exponent. \textbf{(b-inset)} Typical yield strain around blocks that stop at low $x_\varepsilon$ after macroscopic slip (in terms of space and strain history -- $\Delta r$: the number of blocks along the weak layer; $\Delta i$: the previous, current, and next yield strains that the block experiences); the colour bar is chosen such that red(blue) corresponds to yield strain that is higher(lower) than the average. \textbf{(c,d)} Cross-sections of (b-inset) along $\Delta i = 0$ and $\Delta r = 0$ (solid-red); in addition: (solid-black) any block in any state, (dashed-blue) any block after macroscopic slip.}
  \label{fig:si:rho(x):correlate}
\end{figure}

\begin{figure}
  \centering
  \includegraphics[width=.9\textwidth]{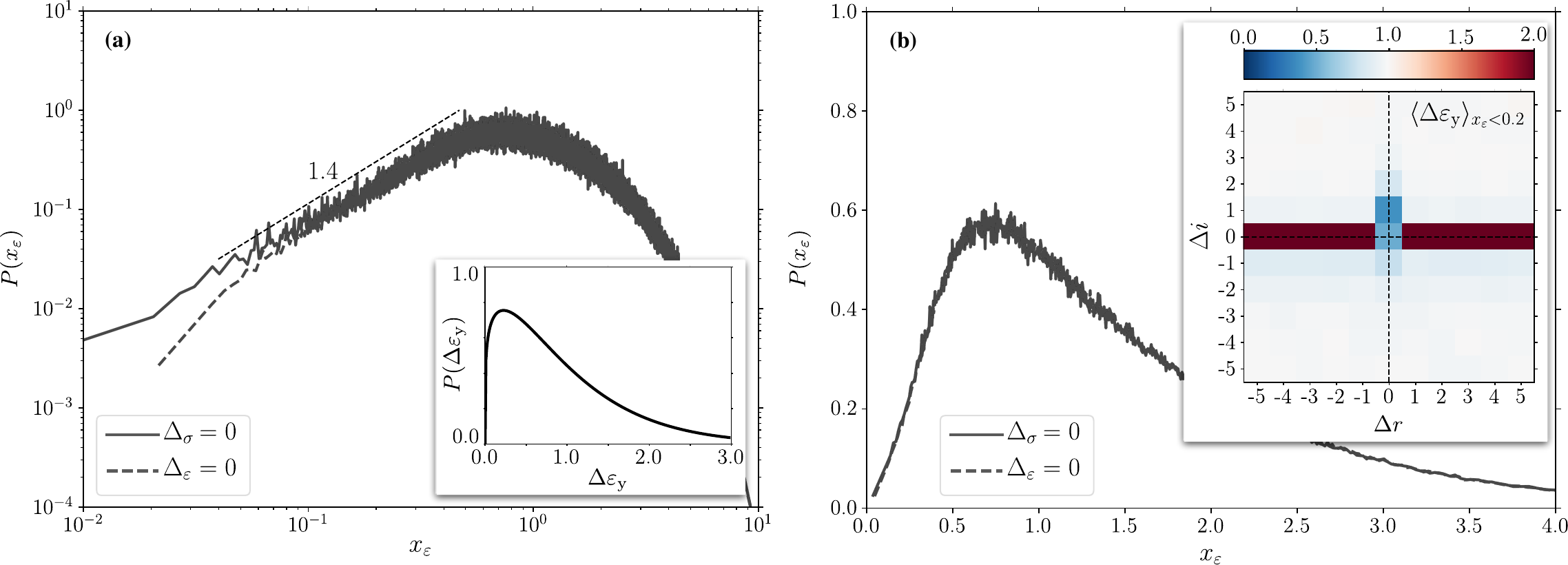}
  \caption{The result in Fig.~\ref{fig:si:rho(x):correlate} based on an ensemble for which the yield strains are drawn from a Weibull distribution characterised by $k = 1.2$ (see Eq.~\eqref{eq:yield-strains}) as shown in \textbf{(a-inset)} (cf.\ Fig.~\ref{fig:si:yield-strains}).}
  \label{fig:distro2:rho(x)}
\end{figure}

We find that those blocks that are close to yielding after macroscopic slip, have strong neighbours while they themselves are weak (their average local yield strain is 46\% lower than the typical one), and also their next yield strain is equally low, see Fig~\ref{fig:si:rho(x):correlate}. This indicates that the block can fail due to mechanical noise, but can also move back because of the low local yield strain. In the dying activity, such block can behave as if there was no inertia, thereby avoiding a gap in $P(x_\sigma)$. This suggests that the presence of pseudo-gap and/or its exponent may depend on microscopic details (as hinted on in \citep{Schwarz2003}). We test this by considering a different distribution of yield strains, representing a different surface roughness, for which it is more likely to find two sequential low yield strains in the local energy landscape. To this end we use the Weibull distribution of Eq.~\eqref{eq:yield-strains} with $k = 1.2$ (instead of $k = 2$). Fig.~\ref{fig:distro2:rho(x)} shows the stability distribution for an ensemble of $40$ realisations of $N = 3^6$. As expected $\theta^\prime$ is decreased with respect to our other results.

\section{Verification of robustness}
\label{sec:robustness}

\subsection{Cutoff}
\label{sec:push_cutoff}

\begin{figure}[htp]
  \centering
  \includegraphics[width=1.\textwidth]{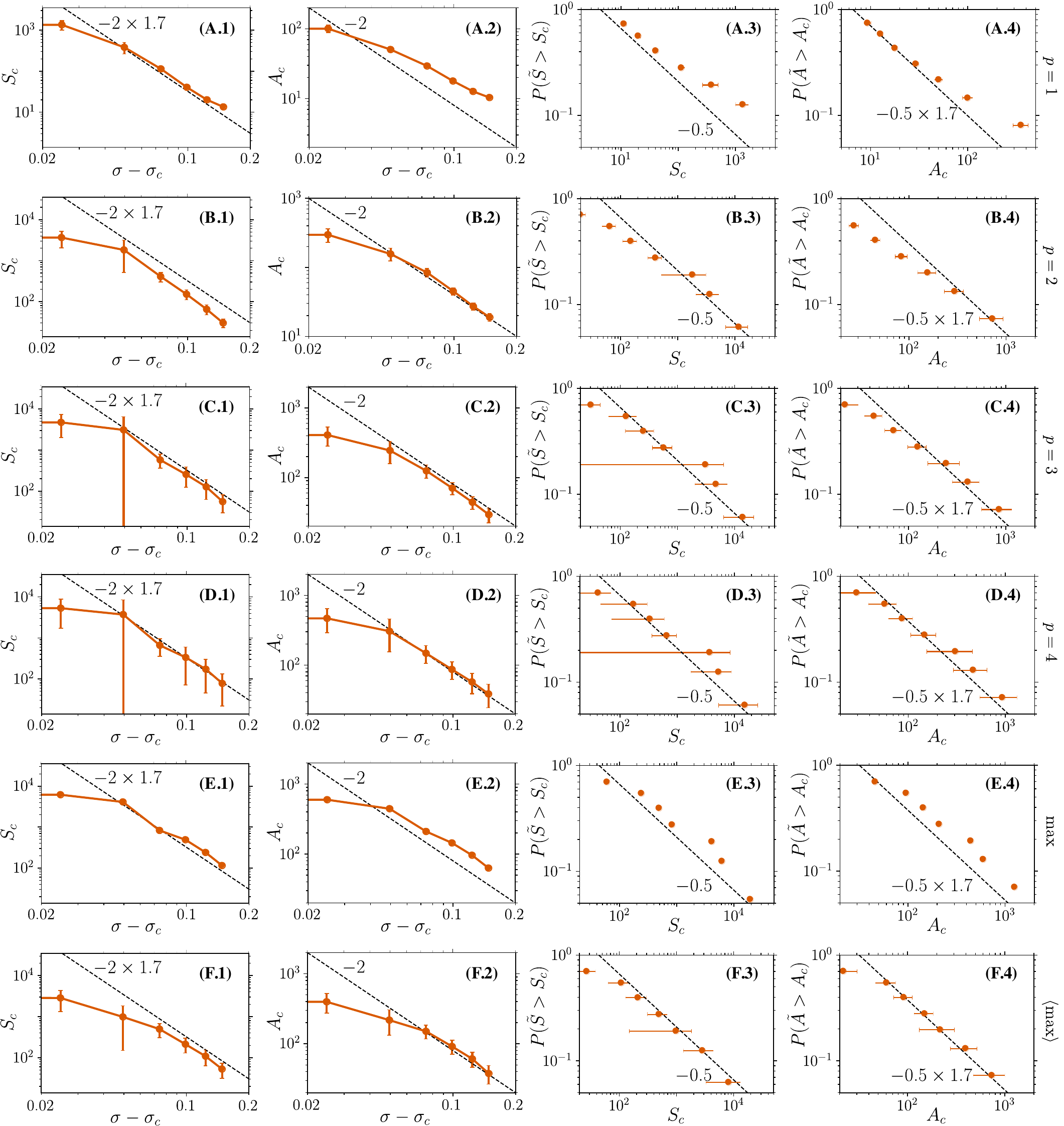}
  \captionof{figure}{Comparison of different measures of cutoff of the size, $S_c$, and area, $A_c$, of avalanches. From left to right: $S_c$ and $A_c$, and the probability that the size or area of any event (avalanche or macroscopic slip) is bigger than $S_c$ or $A_c$ respectively. The different measures, from top to bottom, are defined in Eqs.~(\ref{eq:si:Sc-p}--\ref{eq:si:Sc-max-av}). Note that \textbf{(D.2)} is identical to Fig.~\ref{fig:push}(b) and \textbf{(D.4)} is identical to Fig.~\ref{fig:push}(c). The dashed line marks, in each figure, the power law scaling with the indicated exponent.}
  \label{fig:push_cutoff}
\end{figure}

The proposed critical radius at which an avalanche nucleates macroscopic slip, in Eq.~\eqref{eq:griffith}, can also be expressed in terms of avalanche size using the fractal dimension $d_f$. This corresponds to a critical avalanche size
\begin{equation}
  S_c \sim \Delta_\sigma^{-2 d_f}
  \label{eq:griffith:Sc}
\end{equation}
This scaling is verified in Fig.~\ref{fig:push_cutoff}(D.1). Like for $A_c$, for our measurement we use $S_c \equiv \langle S^{p+1} \rangle / \langle S^{p} \rangle$ with $p = 4$ to be mostly sensitive of the biggest avalanches that did not grow unstable, but the scaling is robust also for different choices of $p$ and for two other protocols, see below. With the measured $S_c$ we can count the fraction of events (avalanches or macroscopic slip) whose size $\tilde{S} > S_c$. Based the power law distribution of avalanche sizes in Eq.~\eqref{eq:rho(S)} we expect:
\begin{equation}
  P(S > S_c)
  \sim \int\limits_{S_c}^\infty P(S) \; d S
  \sim \int\limits_{S_c}^\infty S^{-\tau} \; d S
  \sim S_c^{1-\tau}
  \label{eq:P(Sc)}
\end{equation}
as verified in Fig.~\ref{fig:push_cutoff}(D.3).

Because the measurements of the scaling of the cutoff radius $A_c$ and of the cutoff size $S_c$ are crucial to test our theory, we test the robustness of these measurements. Above we have used
\begin{equation}
  A_c \equiv \langle A^{p+1} \rangle / \langle A^{p} \rangle \qquad
  S_c \equiv \langle S^{p+1} \rangle / \langle S^{p} \rangle
  \label{eq:si:Sc-p}
\end{equation}
with $p = 4$. Here we compare the scaling for different values of $p$, and additionally consider
\begin{equation}
  A_c \equiv \max (A) \qquad
  S_c \equiv \max (S)
  \label{eq:si:Sc-max}
\end{equation}
and
\begin{equation}
  A_c \equiv \langle \max (A) \rangle \qquad
  S_c \equiv \langle \max (S) \rangle
  \label{eq:si:Sc-max-av}
\end{equation}
The latter is the average maximum when all independent measurements are separated in 20 ensembles of measurements. The results are shown in Fig.~\ref{fig:push_cutoff} for all crucial scaling measurements. As observed, all measurements are consistent with our theory, though in particular for $p = 1$ the measurement is polluted by avalanches of small $A$ and $S$.

\subsection{Critical stress \texorpdfstring{$\sigma_c$}{\$sigma\_c\$} and system's size \texorpdfstring{$N$}{\$N\$}}
\label{si:smaller_system}

\begin{figure}[htp]
  \centering
  \includegraphics[width=1.\textwidth]{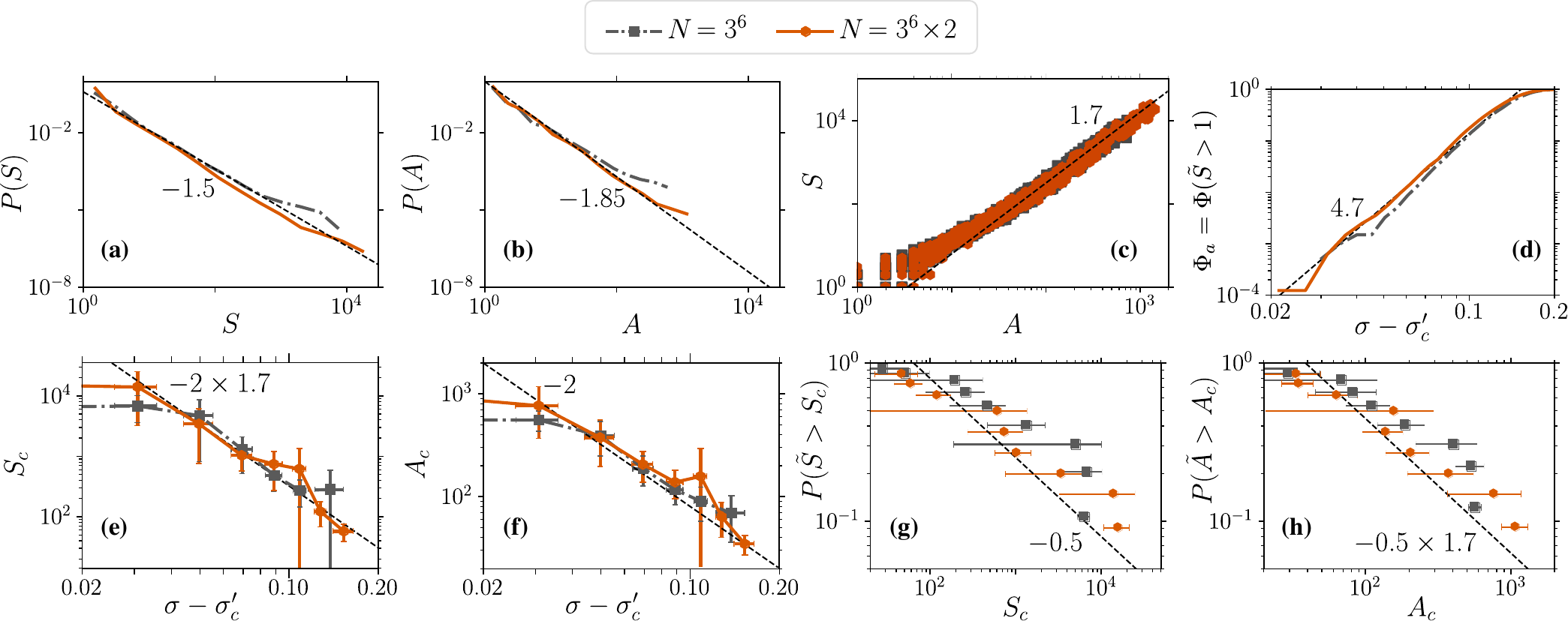}
  \caption{Results using the locally defined critical stress $\sigma_c^\prime$ (defined as the stress after the last macroscopic slip), for two different system sizes. The top row verifies the measured exponents characterising the power law distributions of \textbf{(a)} avalanche sizes ($\tau$, cf.\ Eq.~\eqref{eq:rho(S)} and Fig.~\ref{fig:push}(a)) and \textbf{(b)} avalanche areas (cf.\ Eq.~\eqref{eq:rho(A)} and Fig.~\ref{fig:push}(b)), and \textbf{(c)} the fractal dimension ($d_f$, cf.\ Eq.~\eqref{eq:df} and Fig.~\ref{fig:push}(c))) measured all at $\sigma_c^\prime = 0$; and \textbf{(d)} the fraction of blocks that triggers an avalanche upon increasing the stress by $\Delta_\sigma^\prime = \sigma - \sigma_c^\prime$ ($\theta$, cf.\ Eq.~\eqref{eq:Phi_a} and Fig.~\ref{fig:theta}). The bottom row verifies the prediction scaling for the cutoff of \textbf{(e)} avalanche sizes ($S_c$, cf.\ Eq.~\eqref{eq:griffith:Sc} and Fig.~\ref{fig:push_cutoff}(D.1)) and \textbf{(f)} avalanche areas ($A_c$, cf.\ Eq.~\eqref{eq:griffith} and Fig.~\ref{fig:push_scaling_A}(b)), and corresponding probabilities in terms of \textbf{(g)} avalanche size (cf.\ Eq.~\eqref{eq:P(Sc)} and Fig.~\ref{fig:push_cutoff}(D.3)) and \textbf{(h)} avalanche area ($A_c$, cf.\ Eq.~\eqref{eq:P(Ac)} and Fig.~\ref{fig:push_scaling_A}(c)). The two system sizes that are shown are: $N = 3^6 \times 2$ shown using solid orange lines, and $N = 3^6$ shown using grey dot-dashed lines. The dashed black line marks, in each figure, the power law scaling with the indicated exponent.}
  \label{fig:sigmac}
\end{figure}

In the main text, we measured the scaling of $A_c$, $S_c$, and the number of avalanches (characterised by the exponent and $\theta$) as a function of an increase of stress $\Delta_\sigma = \sigma - \sigma_c$ compared to the ensemble averaged critical stress $\sigma_c \equiv \langle \sigma_c(s) \rangle$ (where $s$ is an index that loops over all macroscopic slips in the ensemble). To verify the robustness of this protocol, we measure the scaling of these quantities as a function of an increase of stress $\Delta_\sigma^\prime \equiv \sigma(s) - \sigma_c(s)$ compared to the local critical stress: the value of stress directly after the last macroscopic slip. The results are fully consistent with the measured scaling relationships in the main text, as shown in Fig.~\ref{fig:sigmac}. Note that, for consistency, we denoted $\sigma - \sigma_c^\prime \equiv \Delta_\sigma^\prime$.

Furthermore, in Fig.~\ref{fig:sigmac} we show all results for two system sizes $N = 3^6 \times 2$ (used in the main text) and $N = 3^6$. Note that even smaller systems do not offer a perspective of validation as those systems do not display a clear separation between avalanches and macroscopic slip, see Appendix~\ref{sec:statistics}. The results of $A_c$ and $S_c$, in Fig.~\ref{fig:sigmac}(e,f), furthermore, emphasise that departure from the predicted scaling of $A_c$ and $S_c$, at small $\sigma - \sigma_c^\prime$, is a finite size effect. In particular, large $A_c$ and $S_c$ are better approximated for the largest system. The departure from scaling happens when $A_c$ is larger than $N / 2$. For such a radius, the periodic repetitions are certainly felt by a propagating avalanche. This causes macroscopic slip to be nucleated sooner, which leads to the observed smaller than predicted $A_c$ and $S_c$.

\section{Required statistics}
\label{sec:statistics}

\begin{figure}[htp]
  \centering
  \includegraphics[width=.4\textwidth]{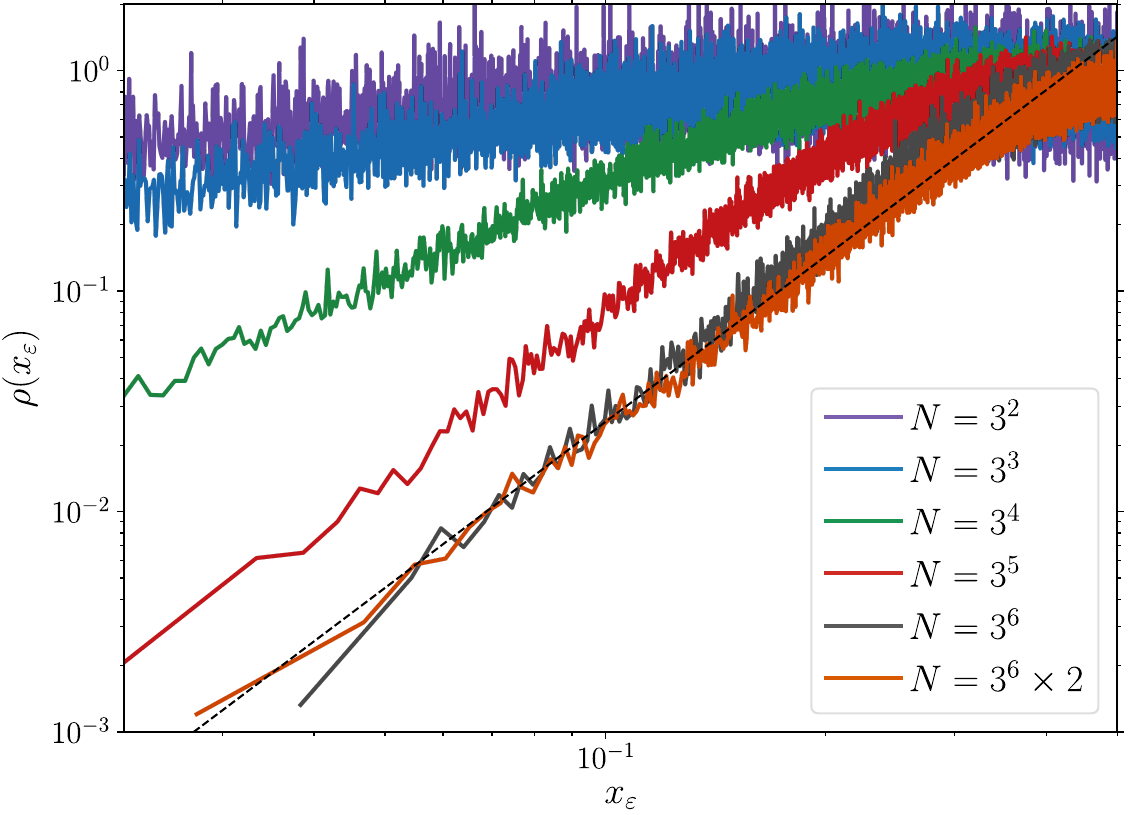}
  \caption{Distribution of the local distance to yielding, $P(x_\varepsilon)$ (see text for definition), at $\sigma = \sigma_c$ for different system sizes $N$ (shown using different colours). The converged exponent $\theta^\prime = 2.5$ is indicated using a dashed line.}
  \label{fig:P(x)_var-N}
\end{figure}

To aid setting up (experimental) validations of our theoretical predictions we list the (statistical) details of our measurements. We emphasise that because each (experimental) protocol comes with its own sources of uncertainties these numbers should be used merely as guideline. We, furthermore, emphasise that the number of avalanches scales with the system size $N$ through the stability exponent $\theta$ as in Eq.~\eqref{eq:n_a}, a fact that can be used to optimise the employed (experimental) protocol.
\begin{itemize}
    \item \emph{System size}: $N$.
    \\
    Throughout the text we use a system size of $N = 3^6 \times 2 \approx 1500$ blocks, whose size is equal to the Larkin length. We find that all our results are robust for a system size of $N = 3^6 \approx 750$ (see Appendix~\ref{sec:robustness}). We find that smaller systems do not display a clear distinction between avalanches and runaway slip events. Such finite size effects also appear in the distribution of the local distance to yielding, $P(x_\varepsilon)$, as shown in Fig.~\ref{fig:P(x)_var-N}.
    \item \emph{Stability distribution}: $P(x_\sigma) \sim (x_\sigma)^\theta$, see Eq.~\eqref{eq:rho(x)}.
    \\
    The result in Fig.~\ref{fig:theta} is based on 8115 avalanches obtained from 2279 steady state stick-slip cycles. Note that using less than 2000 avalanches (or 570 stick-slip cycles) the power law scaling was not obvious.
    \item \emph{Avalanche exponent}: $P(S) \sim S^{-\tau}$, see Eq.~\eqref{eq:rho(S)}, or $P(A) \sim A^{-d_f(\tau-1)-1}$, see Eq.~\eqref{eq:rho(A)}.
    \\
    Fig.~\ref{fig:push}(b) is based on 4513 (triggered) avalanches. Note that we could not extract the correct exponent using less than 500 (triggered) avalanches.
    \\
    Without triggering, the number of avalanches $n_a \sim N (\sigma - \sigma_c)^{\theta + 1}$ (Eq.~\eqref{eq:n_a}). In our model $\theta + 1 \simeq 4.7$ so that avalanches at stresses $\sigma \approx \sigma_c$ are rare. For our model, only 0.2\% of the naturally formed avalanches are in the first bin of Fig.~\ref{fig:push}(b) (i.e.\ at a stress $\sigma \leq \sigma_c + \frac{1}{6}(\sigma_n - \sigma_c)$), one thus needs roughly $6 \cdot 10^4$ steady state stick-slip cycles if triggering is not used.
\end{itemize}

\end{document}